\documentclass[11pt]{elsarticle}
\makeatletter
\def\ps@pprintTitle{%
 \let\@oddhead\@empty
 \let\@evenhead\@empty
 \def\@oddfoot{\centerline{\thepage}}%
 \let\@evenfoot\@oddfoot}
\makeatother

\usepackage{lineno,hyperref}
\modulolinenumbers[255]
\usepackage[top=1.50cm, bottom=1.50cm, left=1.884cm, right=1.884cm, includehead, includefoot, heightrounded]{geometry}

\usepackage{graphicx} 
\usepackage[textfont=bf, font = small]{caption} 
\usepackage{floatrow} 
\usepackage{amssymb}
\usepackage{amsfonts}
\usepackage{amsmath}
\usepackage{color,soul}
\usepackage[symbol]{footmisc}

\usepackage{mathtools}
\DeclarePairedDelimiter\abs{\lvert}{\rvert}

\usepackage{soul,color}
\soulregister\cite7
\soulregister\ref7
\soulregister\pageref7

\begin{document}

\begin{frontmatter}

\title{On the use of high order central difference schemes for differential equation based wall distance computations}

\author{$\text{Hemanth Chandra Vamsi Kakumani}^{a}$}
\author{$\text{Nagabhushana Rao Vadlamani}^{a,}$\footnote{Corresponding author: \texttt{nrv@iitm.ac.in} (Nagabhushana Rao Vadlamani) }}
\author{$\text{Paul Gary Tucker}^{b}$}

\address{$^{a}$\text{Department of Aerospace Engineering, Indian Institute of Technology Madras, Chennai, 600036, India}}
\address{$^{b}$\text{Department of Engineering, University of Cambridge, Cambridge, CB2 1PZ, UK}}


\begin{abstract}
A computationally efficient high-order solver is developed to compute the wall distances by solving the relevant partial differential equations, namely: Eikonal, Hamilton-Jacobi (HJ) and Poisson equations. In contrast to the upwind schemes widely used in the literature, we explore the suitability of high-order central difference schemes (explicit/compact) for the wall-distance computation. While solving the Hamilton-Jacobi equation, the high-order central difference schemes performed approximately $1.4-2.8$ times faster than the upwind schemes with a marginal improvement in the solution accuracy. A new pseudo HJ formulation based on the localized artificial diffusivity (LAD) approach has been proposed. It is demonstrated to predict results with an accuracy comparable to that of the Eikonal equation and the simulations are $\approx$ 1.5 times faster than the baseline HJ solver using upwind schemes. A curvature correction is also incorporated in the HJ equation to correct for the near-wall errors due to concave/convex wall curvatures. We demonstrate the efficacy of the proposed methods on both the steady and unsteady test cases and exploit the unsteady wall-distance solver to estimate the instantaneous shape and burning surface area of a dendrite propellant grain in a solid propellant rocket motor.

\end{abstract}


\end{frontmatter}

\linenumbers

\section{Introduction} \label{intro}

Nearest distance to the solid no-slip wall, referred to as wall-distance in the literature, is often used in the formulation of several eddy-resolving or turbulence modelling strategies. For example, original formulation of the Detached Eddy Simulations (DES) uses wall distance to switch between the modelled and resolved regions of turbulence. Reynolds Averaged Navier Stokes (RANS) \cite{Turb_model} models like Spalart Allmaras, $k-\epsilon$, $k-\omega$ use wall-distance for the near-wall treatment. Such strategies can also be used to alter the dissipation term in the turbulence models or to introduce surface roughness effects by locally altering the length scales of flow using wall-distance \cite{rich,raothesis}. Wall-distances are also used in the peripheral flow applications such as multiphase flows. Accuracy of the wall-distance values and their efficient computation is important in these fluid flow simulations. In addition to Computational Fluid Dynamics (CFD), wall-distances are also encountered in other applications such as Computer Aided Design (CAD) and automated meshing.

There are various search-based algorithms available in the literature to estimate the wall-distances \cite{search1, search2}. However, search algorithms are computationally expensive and suffer from weak scalability, specifically on large meshes and with increasing number of wall faces in the computational domain. Search algorithms also produce inaccurate solutions on non-orthogonal and unstructured grids due to Eucludian and projection based measurement approaches they employ. Search approaches can sometimes produce non-smooth or heavily discontinuous solutions, particularly on low-quality meshes which are non-orthogonal and highly skewed. These inaccuracies can lead to convergence issues of the flow simulations.

On the other hand, differential equation based wall-distance strategies \cite{AR,JL,RS_AST} overcome most of the limitations encountered by the search-based algorithms. These approaches are highly scalable, much more accurate than search-based algorithms and produce smoother solutions in the computational domain. Three classes of differential equation based wall-distance approach include: Eikonal, Hamilton-Jacobi (H-J) and Poisson equations, which will be described in detail in the subsequent section. The formulations of these equations is similar to those of the differential equations governing the flow. Hence, the wall-distance differential equations can be readily implemented into the CFD solvers taking advantage of the existing functions/subroutines of the code that are already written. In addition, the differential equation based approach is demonstrated to be robust even on poor quality grids by Tucker et al. \cite{AR}. Eikonal and H-J equations can be categorized as advection and advection-diffusion equations respectively. Tucker et al. \cite{AR} has proposed upwind schemes to drive the wall-distance equations to a steady state solution. Although upwind schemes are robust and accurate particularly while solving the Eikonal equation, it involves the repeated computation of the wave front direction for every iteration. This results in an increase in the CPU time per iteration and consequently the overall convergence time particularly on unstructured grids. The implementation of upwind scheme is also relatively elaborate in curvilinear coordinates when compared to the standard explicit finite difference or compact schemes.

In the context of high-fidelity eddy resolving simulations like Large Eddy Simulations (LES) or Direct Numerical Simulations (DNS), high-order schemes have been gaining increasing popularity as these schemes can accurately represent the high spatio-temporal frequencies in the transitional/turbulent flows at a substantially lower grid count than the low-order schemes. To accurately capture turbulence, non-dissipative high-order explicit and compact central schemes are preferred over the dissipative up-wind schemes. This leads to the multi-fold objectives of the current work. Firstly, we explore the suitability of using the high-order central difference schemes (explicit and compact schemes up-to 6th order accuracy) to solve the differential equations governing the wall distance. We compare the accuracy and computational efficacy of these schemes with the standard upwind schemes. Subsequently, a modified local artificial diffusivity (LAD) based Hamilton-Jacobi formulation will be introduced, which yields wall-distance estimates as accurate as the Eikonal equation with enhanced stability characteristics. To account for the loss of wall distance accuracy near the regions of high wall curvatures, we introduce a curvature correction to the HJ equation inline with that proposed by Nakanishi \cite{Japanese}. In contrast to the fixed wall test cases typically reported in the literature, we also demonstrate the efficacy of the current formulation on the unsteady moving bodies where the wall-distance changes with time, including the practical test case of the burnback of a propellant grain in a solid rocket motor. It should be noted that the scope of the paper is confined to demonstrating the suitability (both in terms of computational cost and accuracy) of the high-order methods towards computing the wall-distances. Exploring the effects of the aforementioned strategies on the flow-field predictions is beyond the scope of the current study.

Here is the brief outline of the paper. Section \ref{govern} describes the three classes of differential equations governing the wall distances. In Section \ref{numer}, we present the numerical methodology employed for the spatial discretization, filtering and convergence of the baseline and enhanced solvers. The details of the test cases are given in section \ref{casessec}. In section \ref{results}, we present the results for several steady and unsteady test cases using different discretization schemes. We also demonstrate the performance of the proposed modifications (HJ-LAD/Curvature corrections) on both the computational cost and accuracy. We conclude the paper in section \ref{conc} summarizing the key findings of the study.

\section{Governing Equations} \label{govern}

\subsection{Eikonal equation}

The exact governing equation for wall-distance ($\phi$) is the Eikonal equation (Eqn.\ref{ek_eq}), which is a non-linear hyperbolic partial differential equation. It's formulation originates from the fact that the magnitude of wall distance gradient at all the points in the domain equals one \cite{First_EK_paper}. The equation reads as follows.

\begin{equation}\label{ek_eq}
        |\nabla \phi|=1 \hspace{6pt} \quad or \hspace{6pt} (\nabla \phi)^2=1 \hspace{6pt} \quad or \hspace{6pt} \textbf{U}\cdot\nabla \phi=1
\end{equation}
\begin{equation}\label{ek_eq2}
        U_{x}\frac{\partial \phi}{\partial x} +  U_{y}\frac{\partial \phi}{\partial y} +  U_{z}\frac{\partial \phi}{\partial z} =1
\end{equation}

\noindent where $\textbf{U}$ in the Eqn.\ref{ek_eq} is defined as $\nabla \phi$, the Eikonal front propagation velocity. The equation can also be rewritten in its expanded form, Eqn.\ref{ek_eq2}, with $U_{x}$, $U_{y}$ and $U_{z}$ as the components of $\textbf{U}$. Given the non-linearity of this equation, upwind schemes are generally used in the literature. Solving it using non dissipative central schemes is challenging as the high-frequency Gibbs oscillations across the sharp discontinuity of the Eikonal front leads to the divergence of the solution. Dissipation of some form is required to suppress these high frequency oscillations thereby enhancing the stability and ensuring robust convergence. Methods to solve this equation will be described in section \ref{numer}.

\subsection{Poisson equation}

Spalding \cite{Poisson_first} proposed a numerically simplistic Poisson based method to compute wall distances. The governing equation is given by:

\begin{equation}\label{Poiss1}
        \nabla^{2} \phi'=-1
\end{equation}
\begin{equation}\label{Poiss2}
        \phi=\pm \sqrt{\sum_{j=1,3}\left(\frac{\partial \phi'}{\partial x_{j}}\right)^{2}}+\sqrt{\sum_{j=1,3}\left(\frac{\partial \phi'}{\partial x_{j}}\right)^{2}+2 \phi'}
\end{equation}

It should be noted that the approach involves solving for $\phi'$ as the variable in the Poisson equation (Eqn.\ref{Poiss1}). The actual wall distance, $\phi$, is subsequently computed using the auxiliary equation, Eqn.\ref{Poiss2}. This approach only yields accurate wall distances close to the walls. In most of the turbulence modelling strategies, it is sufficient enough to estimate accurate wall-distances close to the walls where the viscous effects are dominant. Nevertheless, certain applications like Computer Aided Design (CAD), computing minimal surfaces, automated meshing \cite{Auto_mesh} etc demand accurate wall distance away from the walls.

\subsection{Hamilton-Jacobi (H-J) equation}\label{hjsection}

Hamilton-Jacobi (HJ) equation, proposed by Tucker et al. \cite{Tucker}, can be interpreted as a hybrid between the Eikonal and Poisson equations. With the advection term on the LHS and a Laplacian term on the RHS, HJ equation takes the general form Eqn \ref{Ham_Jac}.

\begin{equation}\label{Ham_Jac}
       \textbf{U}\cdot\nabla \phi=1+\Gamma\nabla^{2} \phi
\end{equation}

While the hyperbolic nature of advection term helps in algorithmic efficiency, the Laplacian operator on the right provides the required dissipation to suppresses high frequency Gibbs oscillations originating from the advection term. The equation hence offers a good alternative to the Eikonal equation due to its superior stability and robustness. Although the Laplacian operator enhances stability, the accuracy of the wall-distance estimates can deteriorate depending upon the formulation used for the term $\Gamma$.

\subsubsection{Standard formulation for \texorpdfstring{$\Gamma$}{Lg}}

The formulation for $\Gamma$  is generally modelled as a linear function of $\phi$ in the literature \cite{Tucker,AR} as shown in Eqn.\ref{Gamma_form1}. Other non-linear variations (quadratic, cubic, logarithmic etc.) can also be used instead of Eqn.\ref{Gamma_form1}. 

\begin{equation}\label{Gamma_form1}
    \Gamma = \epsilon\phi
\end{equation}

Here, $\epsilon$ is an arbitrary constant typically ranging between 0.1 and 1, chosen based on the mesh and geometry of the simulation. It is evident that $\Gamma$ is zero near the walls where $\phi<<0$ (essentially solving the Eikonal equation) and progressively increases away from the wall. The HJ equation using this formulation is referred to as `standard H-J equation' in the current work. It is worth noting that this Laplacian operator, in the original formulation by Tucker et al. \cite{AR}, has been deliberately tuned for turbulence modelling. Near fine convex features (thin wires) or convex corners, the model exaggerates the wall-distance, $\phi>d_{true}$, to remedy the excessive turbulence dissipation. On the other hand, through $\phi< d_{true}$ at a concave corner, the Laplacian term naturally accounts for the damping effects due to extra walls.

\subsubsection{New Localized Artificial Diffusivity (LAD) based formulation for \texorpdfstring{$\Gamma$}{Lg}} \label{ladnew}

We propose an adaptive formulation for $\Gamma$  based on the Localized Artificial Diffusivity (LAD) approach to improve the accuracy. The formulation is inline with the LAD based shock capturing method proposed by Kawai and Lele \cite{LAD}. At each time step, the dispersion errors originating from the non-linear terms in the governing equation must be damped. We use an adaptive value of $\Gamma$ (Eqn. \ref{Gamma_form2}), instead of the linear variation (Eqn. \ref{Gamma_form1}). It should be noted that $\Gamma$ is proportional to the fourth derivative of $\phi$, which is useful in identifying the localized regions where the dispersion error is dominant. The new formulation introduces a desired amount of dissipation to suppress dispersion thereby reducing the overall error in the final solution.

\begin{equation}\label{Gamma_form2}
    \Gamma=\operatorname{MIN}\left(\epsilon \phi, C\left|\sum_{l=1}^{3}\left[\sum_{m=1}^{3}\left(\frac{\partial \xi_{l}}{\partial x_{m}}\right)^{2}\right]^{2} \frac{\partial^{4} \phi}{\partial \xi_{l}^{4}} \Delta x_{l}^{3}\right|\right)
\end{equation}

\noindent Here, $\xi_{l}$ refers to generalized coordinates $\xi$, $\eta$ and $\zeta$. $C$ is an arbitrary constant which is taken as $0.1$ for all the cases reported in this study. With a 10\% variation in $C$, the global errors are found to vary within 1\%.

\section{Computational Set-up} \label{numer}

In this section, we provide details of the transformed governing equations in curvilinear coordinates, spatial discretization and filtering schemes, and the algorithms employed to solve these equations in the baseline and enhanced codes.

\subsection{Governing equations in general coordinates} \label{geq}

The current solver is written in generalized coordinate system with $\xi$, $\eta$, and $\zeta$ being the coordinate axis directions. Equations \ref{ek_eq2}, \ref{Poiss1} and \ref{Ham_Jac} in Cartesian coordinates are transformed to the curvilinear coordinates as given by the equations \ref{Ek_eq_trans}, \ref{Ham_Jac_trans} and \ref{Poiss_trans} respectively. All the equations were non-dimensionalized with the reference length scale $L_{ref}$ and time scale $\tau$.

\noindent \textbf{Eikonal equation:}

\begin{equation}\label{Ek_eq_trans}
        \hat{U} \frac{\partial \phi}{\partial \xi} + \hat{V} \frac{\partial \phi}{\partial \eta} + \hat{W} \frac{\partial \phi}{\partial \zeta} =1
\end{equation}

\noindent \textbf{Hamilton-Jacobi equation:}

\begin{equation}\label{Ham_Jac_trans}
  \begin{aligned}
   \hat{U} \frac{\partial \phi}{\partial \xi} + \hat{V} \frac{\partial \phi}{\partial \eta} + \hat{W} \frac{\partial \phi}{\partial \zeta} =1+\Gamma \left(\xi_{x} \frac{\partial U_{x}}{\partial \xi}+\eta_{x} \frac{\partial U_{x}}{\partial \eta}+\zeta_{x} \frac{\partial U_{x}}{\partial \zeta} +\xi_{y} \frac{\partial U_{y}}{\partial \xi}+\eta_{y} \frac{\partial U_{y}}{\partial \eta}+\zeta_{y} \frac{\partial U_{y}}{\partial \zeta} + \xi_{z} \frac{\partial U_{z}}{\partial \xi}+\eta_{z} \frac{\partial U_{z}}{\partial \eta}+\zeta_{z} \frac{\partial U_{z}}{\partial \zeta}\right)
  \end{aligned}
\end{equation}

\noindent where, $\hat{U}, \hat{V}$ and $\hat{W}$ are contravariant velocities. $U_{x}$, $U_{y}$ and $U_{z}$ are the components of Eikonal front propagation velocity defined as follows:

\begin{equation*}\label{Contra_vels}
\begin{array}{l}
        \hat{U}=\xi_{x} U_{x}+\xi_{y} U_{y}+\xi_{z} U_{z} \\
        \hat{V}= \eta_{x} U_{x} + \eta_{y} U_{y}+ \eta_{z} U_{z} \\
        \hat{W}=\zeta_{x} U_{x}+ \zeta_{y} U_{y}+ \zeta_{z} U_{z}
\end{array}
\end{equation*}

\begin{equation*}
    U_{x} = \frac{\partial \phi}{\partial x} = \xi_{x}\frac{\partial \phi}{\partial \xi} + \eta_{x}\frac{\partial \phi}{\partial \eta} + \zeta_{x} \frac{\partial \phi}{\partial \zeta}
\end{equation*}
\begin{equation*}
    U_{y} = \frac{\partial \phi}{\partial y} = \xi_{y} \frac{\partial \phi}{\partial \xi} + \eta_{y}\frac{\partial \phi}{\partial \eta} + \zeta_{y} \frac{\partial \phi}{\partial \zeta}
\end{equation*}
\begin{equation*}
    U_{z} = \frac{\partial \phi}{\partial z} = \xi_{z} \frac{\partial \phi}{\partial \xi}  + \eta_{z} \frac{\partial \phi}{\partial \eta} + \zeta_{z} \frac{\partial \phi}{\partial \zeta}
\end{equation*}

\noindent \textbf{Poisson equation:}

\begin{equation}\label{Poiss_trans}
  \begin{aligned}
   \xi_{x} \frac{\partial U_{x}'}{\partial \xi}+\eta_{x} \frac{\partial U_{x}'}{\partial \eta}+\zeta_{x} \frac{\partial U_{x}'}{\partial \zeta} +\xi_{y} \frac{\partial U_{y}'}{\partial \xi}+\eta_{y} \frac{\partial U_{y}'}{\partial \eta}+\zeta_{y} \frac{\partial U_{y}'}{\partial \zeta} + \xi_{z} \frac{\partial U_{z}'}{\partial \xi}+\eta_{z} \frac{\partial U_{z}'}{\partial \eta}+\zeta_{z} \frac{\partial U_{z}'}{\partial \zeta} = -1
  \end{aligned}
\end{equation}
\noindent where, 

\begin{equation*}
    U_{x}' = \xi_{x}\frac{\partial \phi'}{\partial \xi} + \eta_{x}\frac{\partial \phi'}{\partial \eta} + \zeta_{x} \frac{\partial \phi'}{\partial \zeta}
\end{equation*}
\begin{equation*}
    U_{y}' = \xi_{y} \frac{\partial \phi'}{\partial \xi} + \eta_{y}\frac{\partial \phi'}{\partial \eta} + \zeta_{y} \frac{\partial \phi'}{\partial \zeta}
\end{equation*}
\begin{equation*}
    U_{z}' = \xi_{z} \frac{\partial \phi'}{\partial \xi}  + \eta_{z} \frac{\partial \phi'}{\partial \eta} + \zeta_{z} \frac{\partial \phi'}{\partial \zeta}
\end{equation*}

\subsection{Spatial discritization} \label{spatial_disc}

\subsubsection{Baseline solver}
The baseline solver employs standard upwind schemes to estimate the advection analogous terms in the Eikonal and Hamilton-Jacobi equations; and $2^{nd}$ order central difference scheme to estimate the Laplacian term in the H-J and Poisson equations. For example, the first term in the LHS of Eikonal equation (Eqn.\ref{Ek_eq_trans}), $\hat{U}\phi_{x}$, is discretized using the up-wind scheme as follows:

\begin{equation}\label{UW1}
        \hat{U} \phi_{x} = 0.5 (\hat{U} + |\hat{U}|) B + 0.5 (\hat{U} - |\hat{U}|) F
\end{equation}
\noindent where,

\begin{equation}\label{UW2}
     B = \frac{\phi_{i}-\phi_{i-1}}{\Delta \xi_{i-1}} \hspace{0.3cm} and \hspace{0.3cm}  F = \frac{\phi_{i+1}-\phi_{i}}{\Delta \xi_{i+1}}
\end{equation}
\noindent Here, `$i$' corresponds to the grid point index in the $\xi$-direction. The terms $\frac{\partial \phi}{\partial \xi}$, $\frac{\partial \phi}{\partial \eta}$ and $\frac{\partial \phi}{\partial \zeta}$, which are required to compute $\hat{U}$, are estimated as follows:

\begin{equation}\label{UW3}
    \frac{\partial \phi}{\partial \xi} \approx n_{i-1} F +n_{i+1}B
\end{equation}
\begin{equation}\label{UW4}
    n_{i-1}=0.25(1+\operatorname{SIGN}(1, F+B)\left(1+\operatorname{SIGN}\left(1,B\right)\right)
\end{equation}\begin{equation}\label{UW5}
    n_{i+1}=0.25(1-\operatorname{SIGN}(1, F +B)\left(1-\operatorname{SIGN}\left(1,F\right)\right)
\end{equation}

Similar formulation is employed to estimate the metric terms and the derivatives of $\phi$ in other coordinate directions, $\frac{\partial \phi}{\partial \eta}$ and $\frac{\partial \phi}{\partial \zeta}$.

\subsubsection{Enhanced solver}

In the enhanced solver, high-order central difference schemes namely explicit second order (E2), explicit 4th order (E4), compact 4th order (C4) and compact 6th order (C6) schemes are used to discretize all the spatial derivatives and metric terms in the governing equations. Implementation of these schemes was straightforward as most of the subroutines were reused from the in-house high-fidelity fluid flow solver COMPSQUARE \cite{COMP_SQ1,COMP_SQ2,COMP_SQ3}. Following \cite{visbal}, the first-order derivative of a variable $\phi$ in the interior domain is computed as follows:
\begin{equation}
	\alpha \phi_{i-1}^{\prime}+\phi_{i}^{\prime}+\alpha \phi_{i+1}^{\prime}=b \frac{\phi_{i+2}-\phi_{i-2}}{4 \Delta \xi}+a \frac{\phi_{i+1}-\phi_{i-1}}{2 \Delta \xi}
\end{equation}
Here, $\phi'$ represents the derivative of $\phi$ in $\xi$ direction. While the constant $\alpha$ is zero for explicit schemes (E2 and E4), it is non-zero for compact schemes (C4 and C6). Values of the scheme-specific coefficients `a' and `b' and the formulation of derivatives at the boundaries can be found in Ref.\cite{visbal}.

\subsection{Solution convergence}

For steady cases, a pseudo time derivative, $\frac{\partial \phi}{\partial t^{*}}$, has been added to the LHS of all the three differential equations (Eqn.\ref{USeq1} \ref{USeq3} \& \ref{USeq4}). An explicit 4-stage 4th order Runge-Kutta (RK) \cite{visbal} is used to drive the solution to a steady state in pseudo-time, $t^{*}$. In addition, local-time stepping is used to accelerate the convergence.

\begin{equation}\label{USeq1}
        \frac{\partial \phi}{\partial t^{*}} + \textbf{U} \cdot \nabla \phi=1
\end{equation}

\begin{equation}\label{USeq3}
        \frac{\partial \phi}{\partial t^{*}} + \textbf{U} \cdot \nabla \phi=1+(\nu\phi) \nabla^{2} \phi
\end{equation}

\begin{equation}\label{USeq4}
        \frac{\partial \phi}{\partial t^{*}} -1 = \nabla^{2} \phi
\end{equation}

\noindent Following convergence criterion is used to arrive at steady state solution:

\begin{equation}\label{Con-criter}
       \abs*{\frac{\phi-\phi_{o}}{L_{ref}}}_{max}  \leq 10^{-5}
\end{equation}

\noindent Here, $\phi$ and $\phi_{o}$ correspond to the wall distance values in the current and previous pseudo time iterations. The solution is converged until the maximum change in the wall distance value during a pseudo time step is below $10^{-5}$ in the entire domain. For the unsteady cases, where the body(s) or the wall(s) move w.r.t time, the boundary conditions in the computational domain are updated at every physical time step and $\phi$ is repeatedly computed. Between two physical time steps, pseudo iterations with local time-stepping are used to drive the solution to a steady state.

\subsection{Filtering scheme}

High frequency dispersion errors arising due to the non-dissipative nature of the high-order central schemes (E2, E4, C4 and C6), and the non-linear nature of the governing equations, can lead to convergence issues. To counter this, an implicit 10th order spatial filter is used to stabilize the solution along all the directions after each pseudo time step. Following \cite{visbal}, the implicit filtering operation of a variable $\Phi$ along a direction $i$ is given as:

\begin{equation}
\alpha_{f} \hat{\phi}_{i-1}+\hat{\phi}_{i}+\alpha_{f} \hat{\phi}_{i+1}=\sum_{n=0}^{n=N} \frac{a_{n}}{2}\left(\phi_{i+n}+\phi_{i-n}\right)
\end{equation}

where $\phi$ and $\hat{\phi}$ are unfiltered and filtered functional values respectively and $\alpha_{f}$ is the filtering coefficient. $2N$ is the order of filter which uses a $2N+1$ point stencil. A 10th order filter ($N$=5) is used in the current work at interior points and $\alpha_{f}$ in the current work is varied between 0.48 and 0.495 for different test cases. The values of filter coefficients, $a_{n}$, at the interior and boundary points are taken from Visbal et al. \cite{visbal}. Following their strategy, the filtering order is gradually reduced up to second order at the boundaries.

\subsection{Boundary Conditions and Initialization}

\begin{figure}[h!]
\begin{center}
\includegraphics[width=130mm]{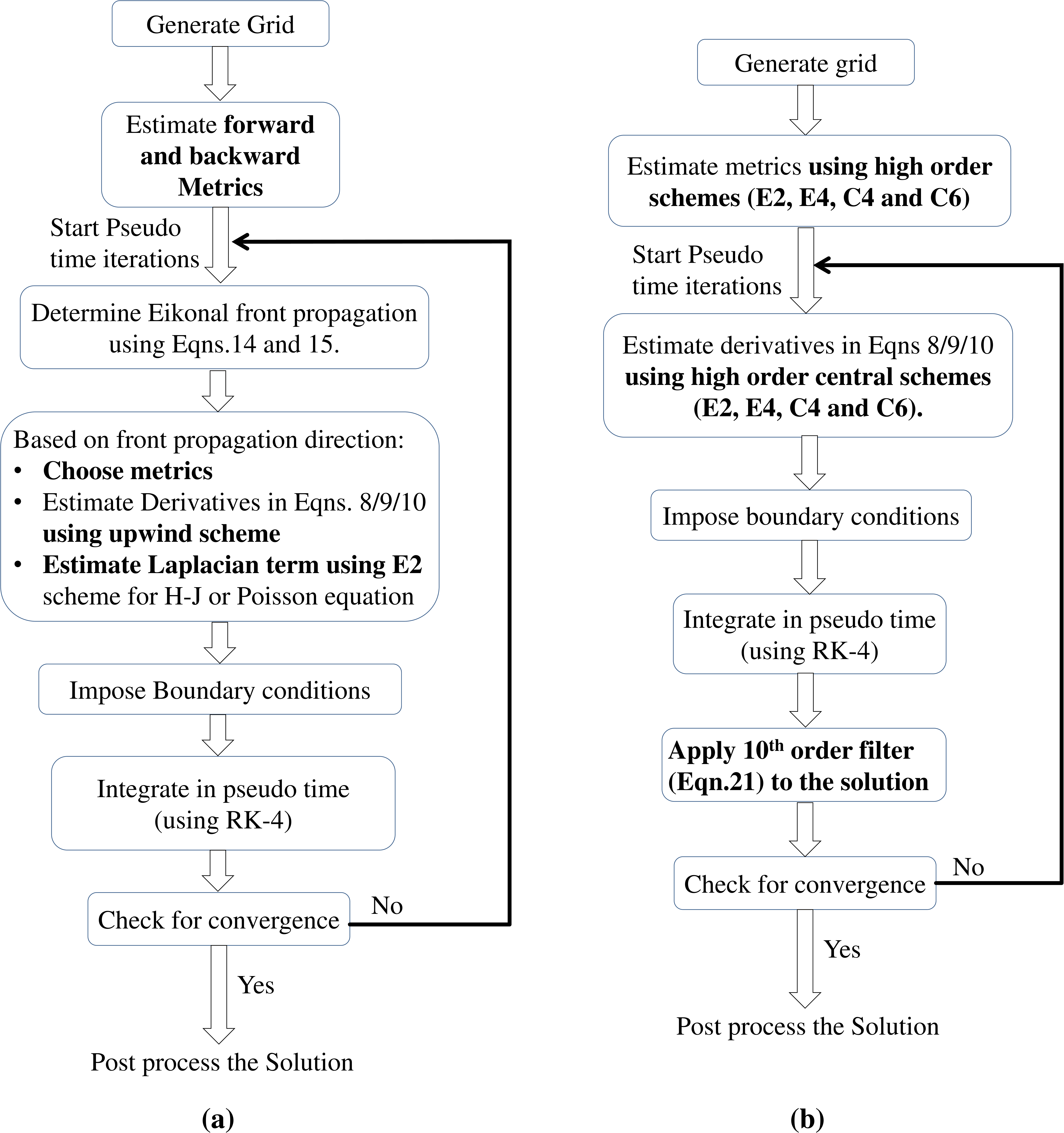}
\end{center}
\caption{Algorithm flow chart for (a) Baseline UW solver and (b) Enhanced high-order solver}
\label{flow_chart} 
\end{figure}

Dirichlet boundary condition is imposed for the grid points on the wall, where the value of the wall distance is set to zero. Wall distance $\phi$, is also set to zero at all the grid points lying within the volume of the solid body/bodies.
\begin{equation}\label{eq7}
        \phi_{wall} = 0
\end{equation}
Neumann boundary condition is imposed at the far-field boundaries, where the gradient of wall distance is set to zero.
\begin{equation}\label{eq8}
        \frac{\partial \phi}{\partial n}=0
\end{equation}
Here, $n$ is the direction normal to the boundary. The wall distance values are initialized to zero within the domain before starting the simulation. Figure \ref{flow_chart} shows the flow charts with the details of the algorithm for the baseline and enhanced solver.

\section{Test cases} \label{casessec}

The implementation of both the baseline and enhanced solvers has been tested on several steady and unsteady test cases listed in Table-\ref{Cases}. Cases 1 and 2 are used to validate the baseline solver of which Case-2, chosen from \cite{RS_AST}, demonstrates the capability of the solver in estimating the wall-distance field around the complex bodies within the computational domain. Cases 3, 4 and 5 are used to validate the enhanced solver and compare the results against the baseline solver. Case-6 is solved to examine the effect of curvature on the accuracy of wall distances computed using H-J approach. Results using standard H-J and standard H-J based on the curvature corrections introduced by Nakanashi \cite{Japanese} are compared. We also discuss the modifications to the Nakanashi's \cite{Japanese} approach, which have improved the accuracy further. Cases 7 and 8 respectively simulate a piston-cylinder arrangement and a cube bouncing between parallel walls. These test cases demonstrate the efficacy of the unsteady solver where the wall-distances between the surfaces vary with time. In addition, we have also simulated the wall distance evolution inside a solid rocket motor with a progressively burning propellant grain.

\begin{table}[h!]
\begin{center}
\includegraphics[width=175mm]{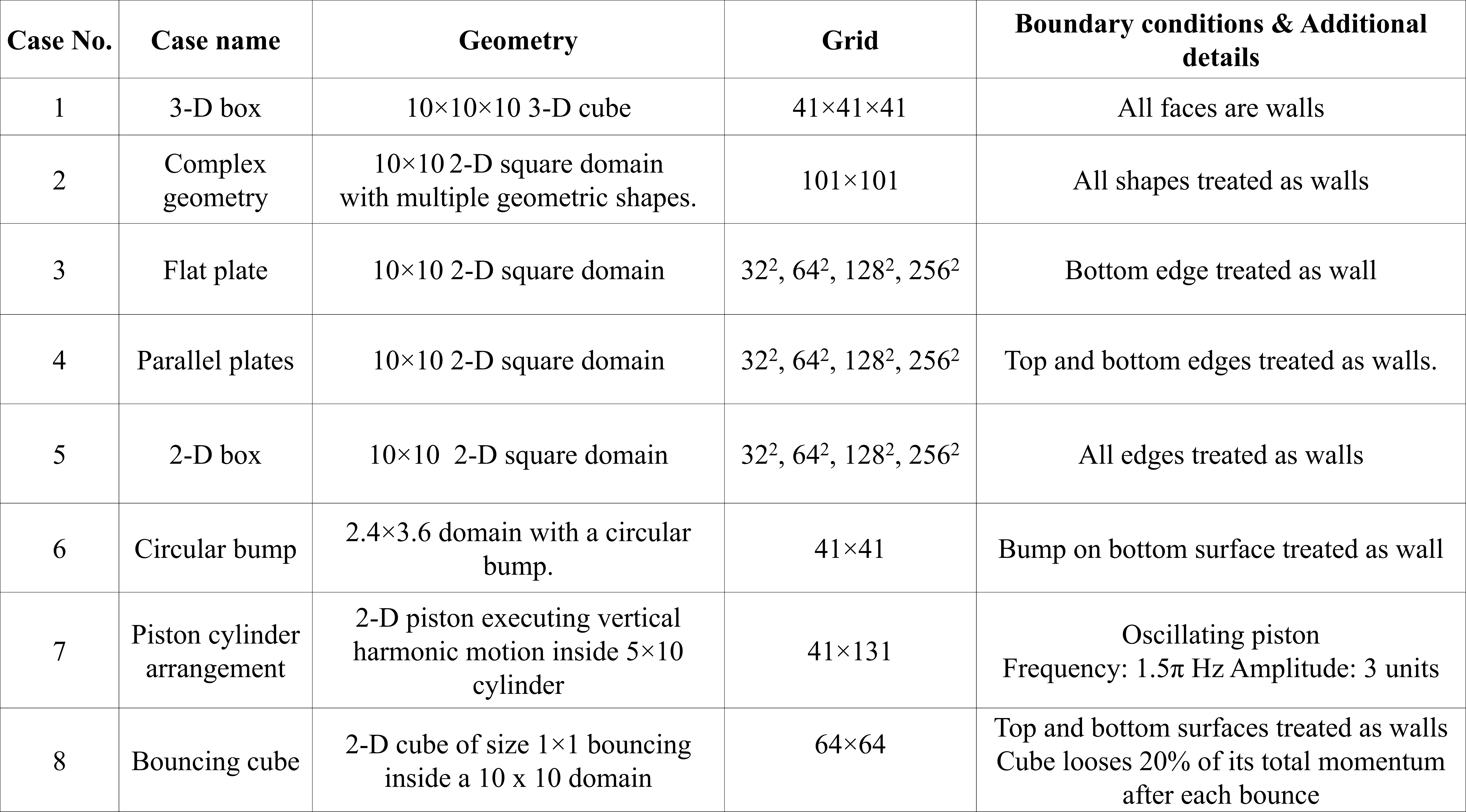}
\end{center}
\caption{Case set-up, boundary conditions and grid details for various canonical test cases.}
\label{Cases} 
\end{table}

\section{Results} \label{results}

In this section the baseline solver which employs first order up-wind scheme is validated on cases 1 and 2 (3-D box and Complex geometries). Subsequently, the enhanced solver is validated on rest of the test cases. For all the test cases, we compare the wall-distance and the error field computed from enhanced solver against the exact results and those obtained using the baseline solver. The exact solution was found analytically based on the case geometry, as most of the test cases used in this work are canonical in nature. Computational time (CPU time) required by each of these methods are also compared. 

\subsection{Validation of the baseline solver}

\begin{figure}[h!]
\begin{center}
\includegraphics[width=140mm]{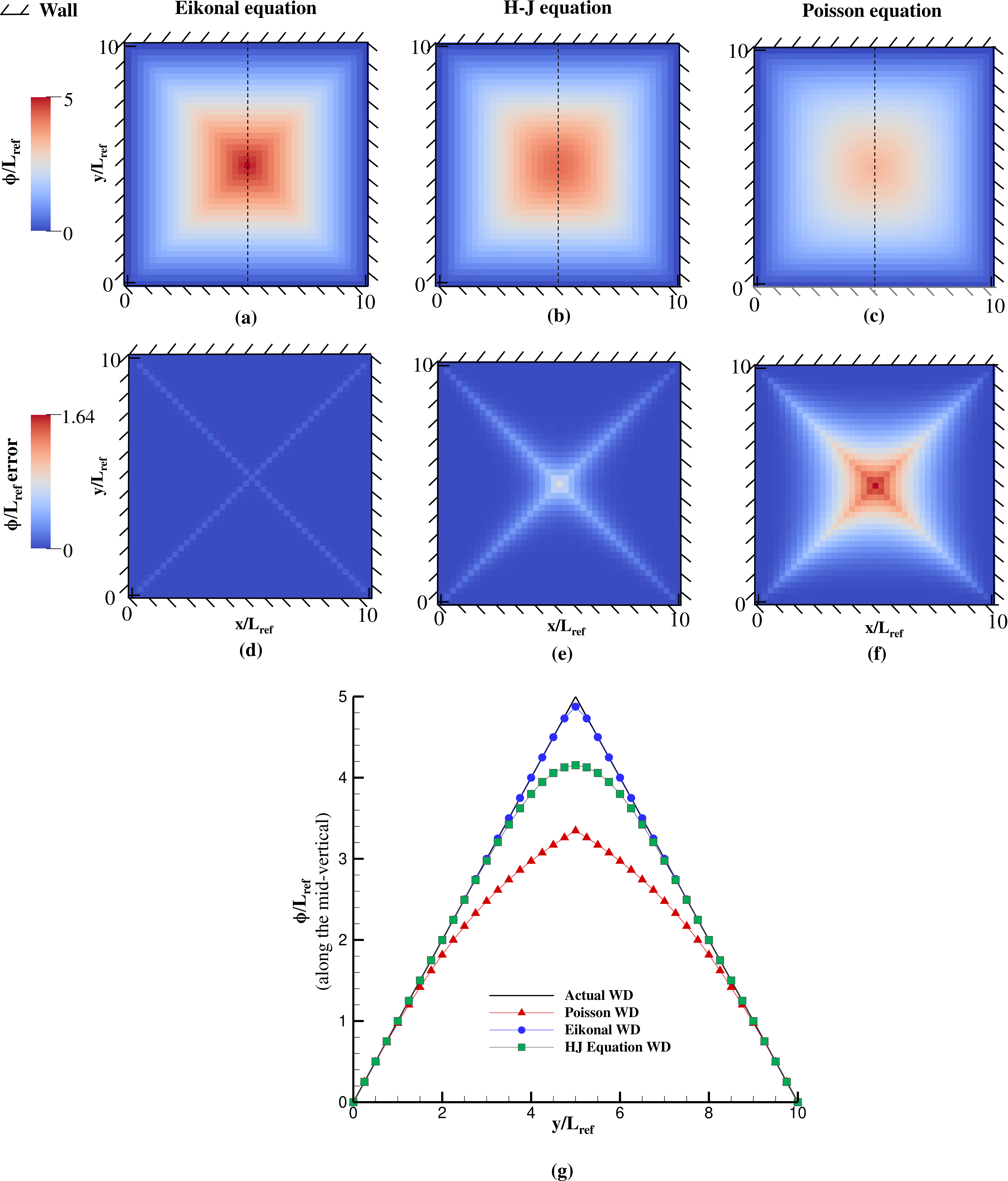}
\end{center}
\caption{Wall distance field (a,b,c) and the corresponding error field (d,e,f) for 3D box case (case-1) obtained by solving Eikonal, Hamilton-Jacobi equation and Poisson equation with the baseline solver (g) Comparison of wall distance along the mid-vertical line with the exact solution.}
\label{baseline1} 
\end{figure}

\begin{figure}[t!]
\begin{center}
\includegraphics[width=160mm]{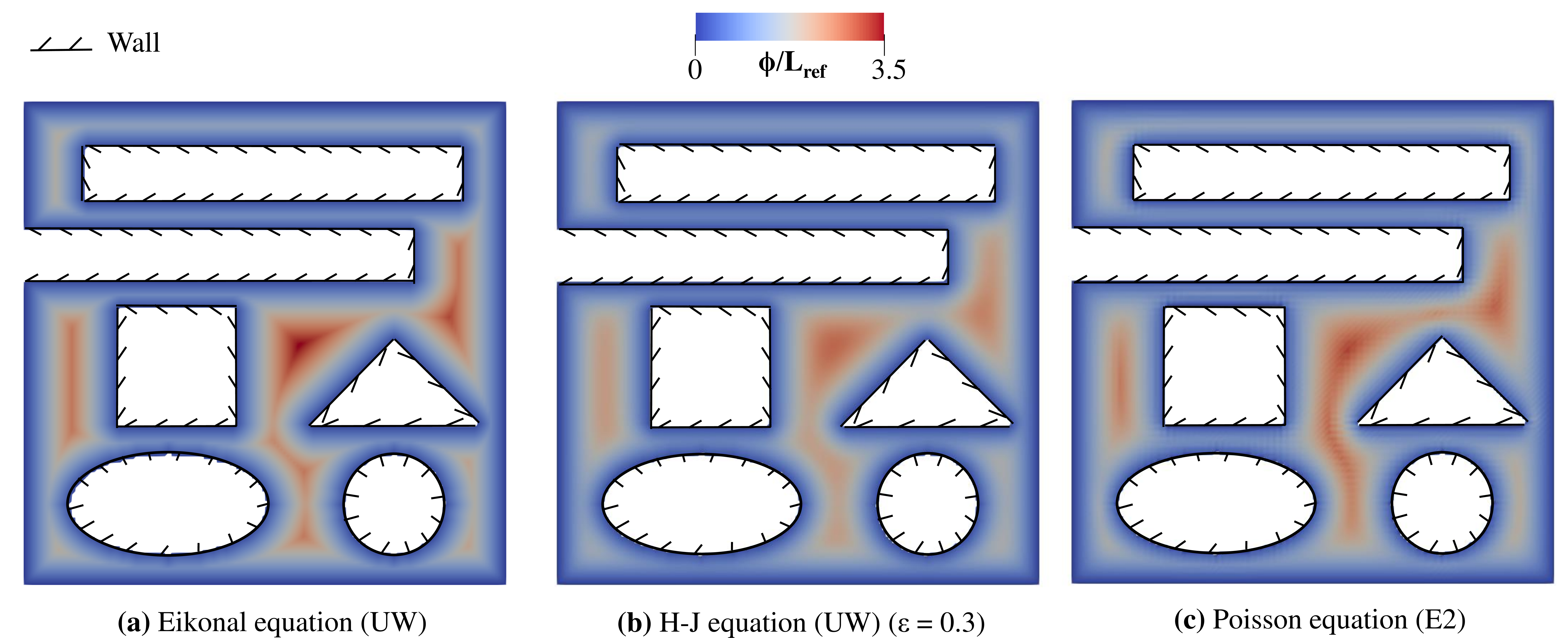}
\end{center}
\caption{Wall distance field obtained using (a) Eikonal, (b) Hamilton-Jacobi and (c) Poisson equations for complex geometries case (case-2).}
\label{baseline2} 
\end{figure}

Figures \ref{baseline1}(a,b,c) depict the wall-distance field computed within a 3D box (case-1) on Z = 5 plane. Corresponding errors w.r.t the exact wall-distance are shown in Figures \ref{baseline1}(d,e,f). We compare the solutions predicted using the Eikonal, H-J and Poisson based approaches. For clarity, Figure \ref{baseline1}(g) also shows the wall-distance on a vertical line marked in Figure \ref{baseline1}(a). The errors are negligible close to the walls for all the approaches. However, it is evident that the highest errors are recorded at the centre of the box and across the diagonals, particularly with the Poisson equation. On the other hand, the predictions of both the Eikonal and H-J are much more accurate, with the Eikonal solution being the most accurate throughout the domain. As noted in section \ref{hjsection}, the loss of accuracy away from the walls with H-J is attributed to the Laplacian term using the standard $\Gamma$ formulation. 
\noindent Figure \ref{baseline2} also compares the computed wall distance contours for a test case with complex geometries embedded in a square domain (case-2). The wall-distance field obtained is in agreement with the results reported in \cite{RS_AST} further validating the framework. It also demonstrates the robustness and performance of the current solver in handling geometries with complex shapes.

\subsection{Effect of high order schemes and LAD formulation on the solution of Hamilton Jacobi Equation}\label{hjsec}



\begin{figure}[h!]
\begin{center}
\includegraphics[scale=0.55]{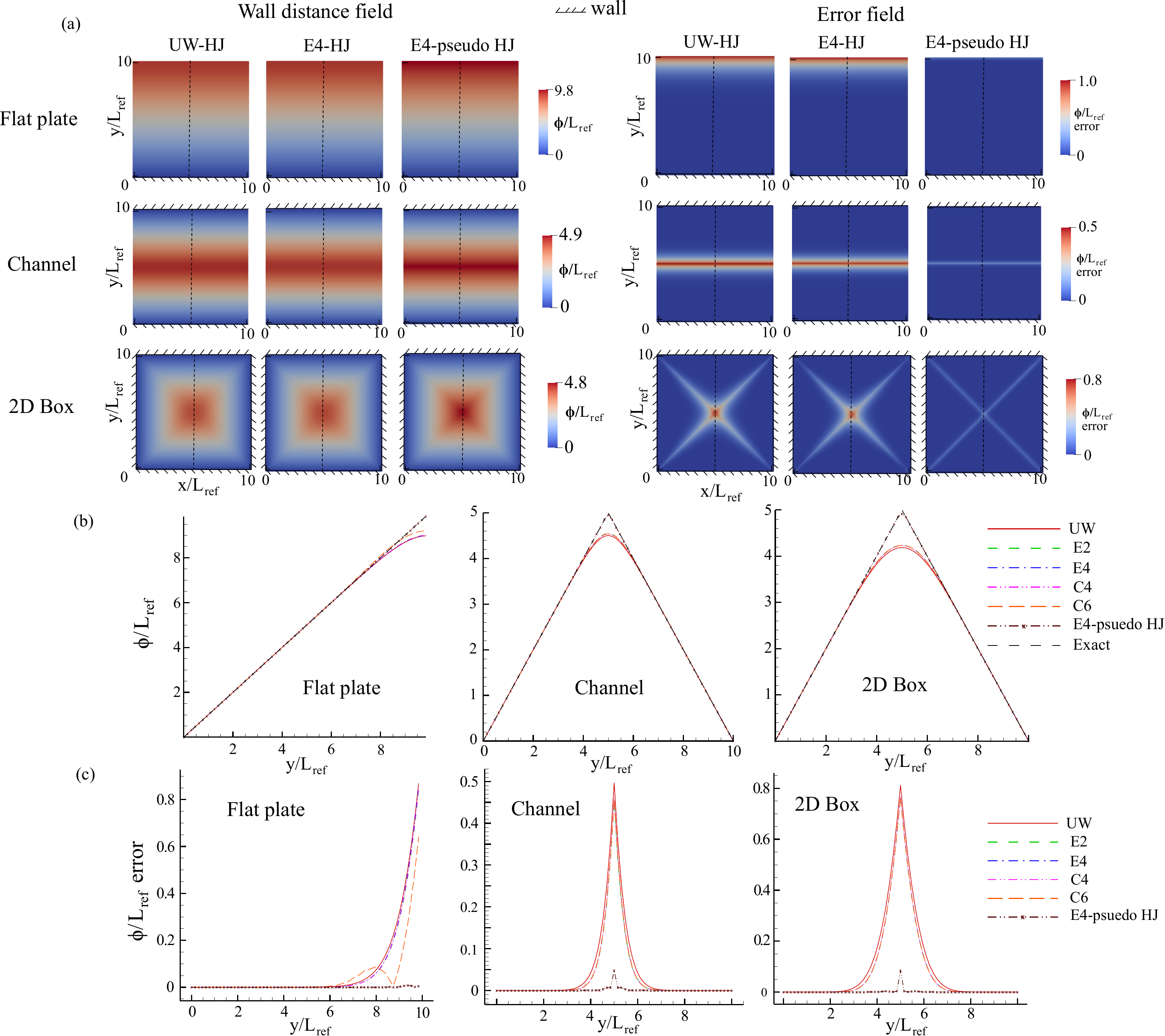}
\end{center}
\caption{(a) Contours of wall distance field and absolute wall distance error. Variation of (b) wall distance and (c) absolute wall distance error along mid vertical line (marked with dashed lines in contour plots) for flat plate, channel and 2D box cases.}
\label{HJcases} 
\end{figure}

Having validated the baseline solver using the upwind schemes, we have solved the same set of differential equations with the high-order central schemes inbuilt in the in-house solver COMPSQUARE. As demonstrated in the flow chart of Figure \ref{flow_chart}(b), the wall-distance solver has been enhanced to exploit the high order explicit/compact discretization schemes (E2,E4,C4,C6) with a tenth order filter (F10). The efficacy of the new LAD based formulation for $\Gamma$ (Eqn. \ref{Gamma_form2}) in the H-J equation has also been examined. The performance of the enhanced solver, in terms of accuracy and speed, is reported in this section.

\begin{figure}[h!]
\begin{center}
\includegraphics[scale=0.85]{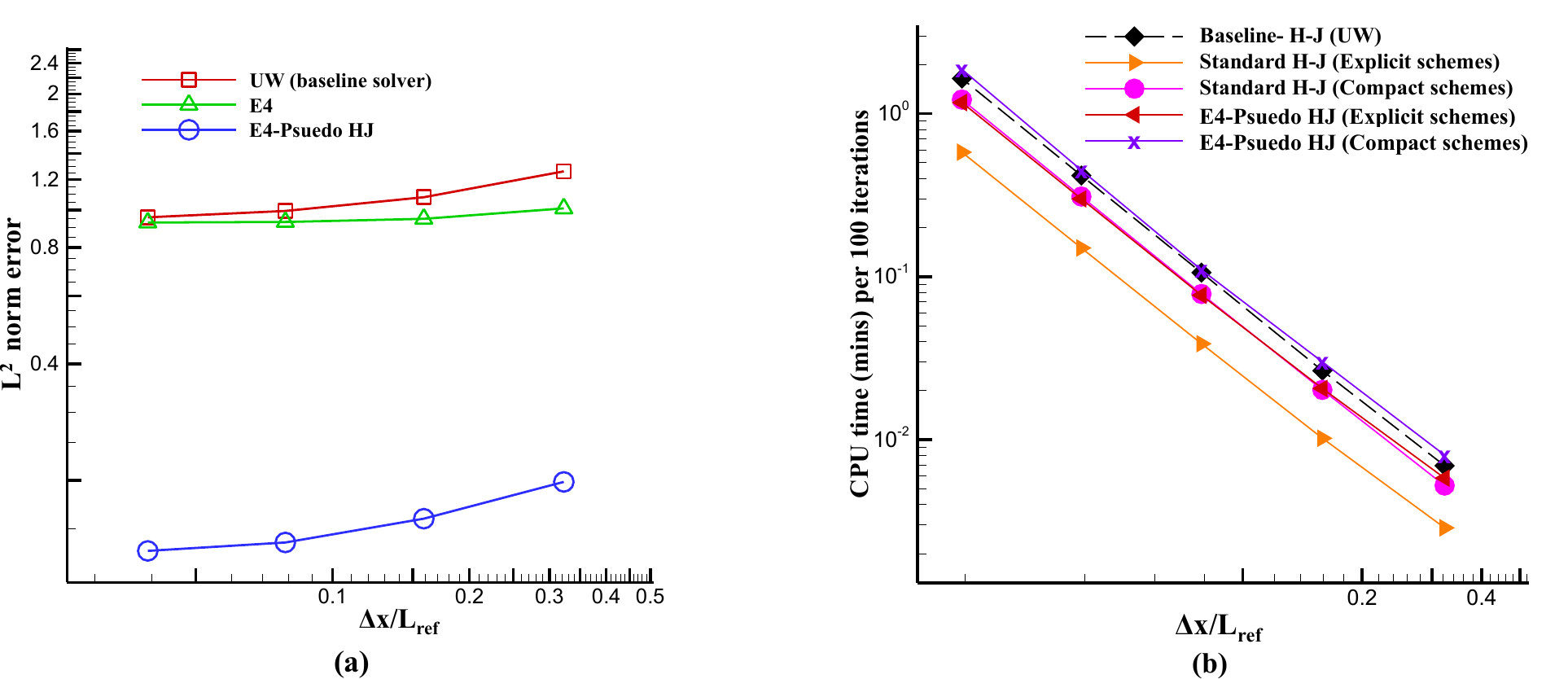}
\end{center}
\caption{(a) $L^2$ norm error with varying grid resolutions and (b) CPU time taken per 100 iterations with different schemes for channel case.}
\label{accspeed} 
\end{figure}

Figure \ref{HJcases}(a) present the contours of wall distance ($\phi/L_{ref}$) and the corresponding errors for the canonical test cases of `Flat plate', `channel' and `2-D box' respectively. For clarity, we have also included line plots of the wall distance (Figure \ref{HJcases}(b)) and the errors (Figure \ref{HJcases}(c)) along the vertical line (shown with dotted lines in the corresponding figures). From the error contours and line plots, it is evident that the predictions of the Pseudo HJ approach with LAD based formulation are in close agreement with the exact solution for all the three test cases than the standard HJ. Selective amount of dissipation added using the LAD formulation (Eqn.\ref{Ham_Jac} and \ref{Gamma_form2}) has remarkably improved the accuracy while maintaining the stability. On the other hand, a marginal improvement in the accuracy is observed with high order schemes (E2, E4, C4 and C6) when compared to the UW scheme. To quantify the improvements in accuracy, simulations are carried out at different grid resolutions for the channel case. Subsequently, the global $L^{2}$ norm is estimated as follows:
\begin{equation}\label{l2norm}
L^{2}=\left(\Delta x \Delta y \sum_{i} \sum_{j}\left|\phi - \phi_{exact}\right|^{2}\right)^{1 / 2}    
\end{equation}
Figure \ref{accspeed}(a) compares the $L^2$-norm for the channel case at different grid resolutions. As expected, the benefit of high order schemes is much more noticeable on a coarser grid than on finer grids. In addition, when compared to the standard HJ formulation, the $L^2$-norm with LAD based formulation is almost an order of magnitude lower.

Figure \ref{accspeed}(b) compares the computational time per 100 iterations for the baseline and enhanced H-J solvers. All the simulations are carried out on a single processor (Dell Optiplex 3070 tower using 3 GHz Intel Core i7-9700 CPU with 32 GB/2666 MHz DDR4 memory). Unlike central difference schemes, upwind schemes involve the repeated computation of the wave front direction at every iteration. Hence, the explicit schemes (E2 and E4) employing the standard and LAD based H-J formulation, are $\approx 2.8\times$ and $\approx 1.5\times$ faster than the baseline solver respectively. LAD based H-J formulation involves computing $\Gamma$ during each iteration and are hence slower than the standard H-J formulation using explicit schemes. Interestingly, the speed of the baseline solver using UW schemes and the enhanced solver using compact schemes for the LAD based H-J equation are almost similar. To summarize, the advantage of using the high-order central difference schemes with filtering over the UW schemes, both in terms of the speed up and accuracy, can be appreciated while solving the H-J equation.



\pagebreak

\subsection{Effect of High order schemes on the solution of Eikonal equation}\label{eiksec}



\begin{figure}[h!]
\begin{center}
\includegraphics[scale=0.85]{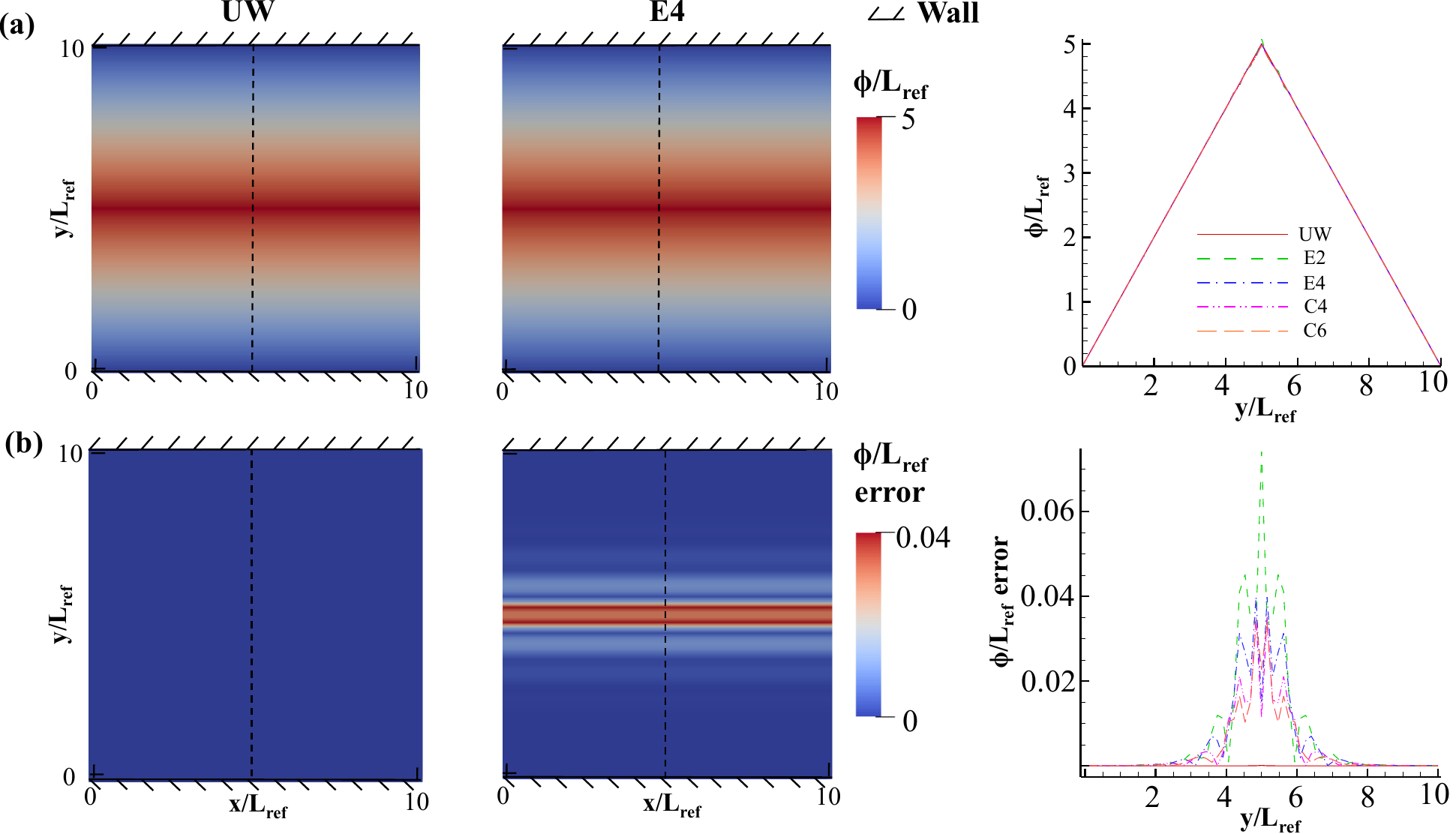}
\end{center}
\caption{Contours of (a) Wall distance field and (b) absolute wall distance error field for the channel case predicted by solving Eikonal equation with UW and E4 schemes. Corresponding line plots along mid-vertical line are also shown}
\label{Eikonal2} 
\end{figure}

\begin{figure}[h!]
\begin{center}
\includegraphics[scale=0.9]{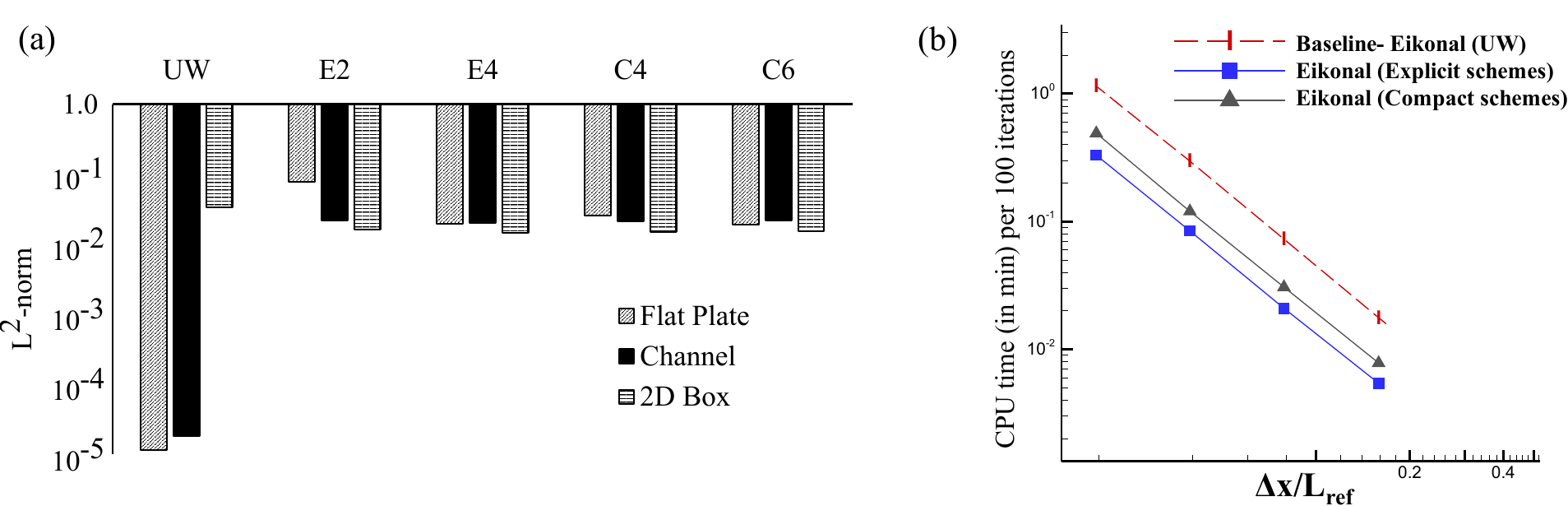}
\end{center}
\caption{Comparison of (a) L2 norm for flat plate, channel and 2D box cases using different schemes (b) CPU time taken per 100 iterations by the baseline and enhanced solver for 2D box case.}
\label{Eikonal4} 
\end{figure}

Inline with the analysis carried out for the H-J equation, we have tested the efficacy of high-order schemes to solve the Eikonal equation. Performance of various schemes is tested on the `channel' case. Figures \ref{Eikonal2}(a,b) show the wall distance field and the error contours in addition to the line plots along the domain center. It is evident that the wall distance predictions with UW and central difference schemes are in close agreement; and any improvement in accuracy due to high order is marginal. In fact, when compared to the H-J equation, the Eikonal equation is hyperbolic and high-order schemes are susceptible to stability issues due to the lack of the Laplacian term. Dispersion errors are hence evident at the centre of the channel (although away from the walls) where the fronts propagating from both the walls meet. Similar conclusion can be drawn from the bar charts shown in Figure \ref{Eikonal4}(a), where we compare the $L^{2}$ norm error of the baseline (UW) and enhanced solvers for three different test cases. The errors for the flat plate and channel cases are lower using the UW scheme when compared to the high-order schemes. For the 2D box case, the errors corresponding to UW and central-schemes are comparable and are predominant across the diagonals (similar to the errors shown in Figure \ref{HJcases}). The results suggest that although the dispersion due to Gibbs phenomenon can be partially suppressed with filtering, the errors due to central schemes while solving the Eikonal equation still persist in the regions of steep gradients where the wave fronts meet. A clear advantage of central schemes is notable in Figure \ref{Eikonal4}(b), where the computational time required for 100 pseudo iterations are compared for different schemes for the `2-D box' case. Due to the lack of the Laplacian term, computational cost associated with solving the Eikonal equation is relatively lower than H-J. The trends are similar to those observed in Fig.\ref{accspeed}(b) with the compact schemes (C4/C6) and Explicit schemes (E2/E4) being $\approx$ 2.3 to 3.5 times faster than the UW solver respectively.

\begin{figure}[h!]
\begin{center}
\includegraphics[width=155mm]{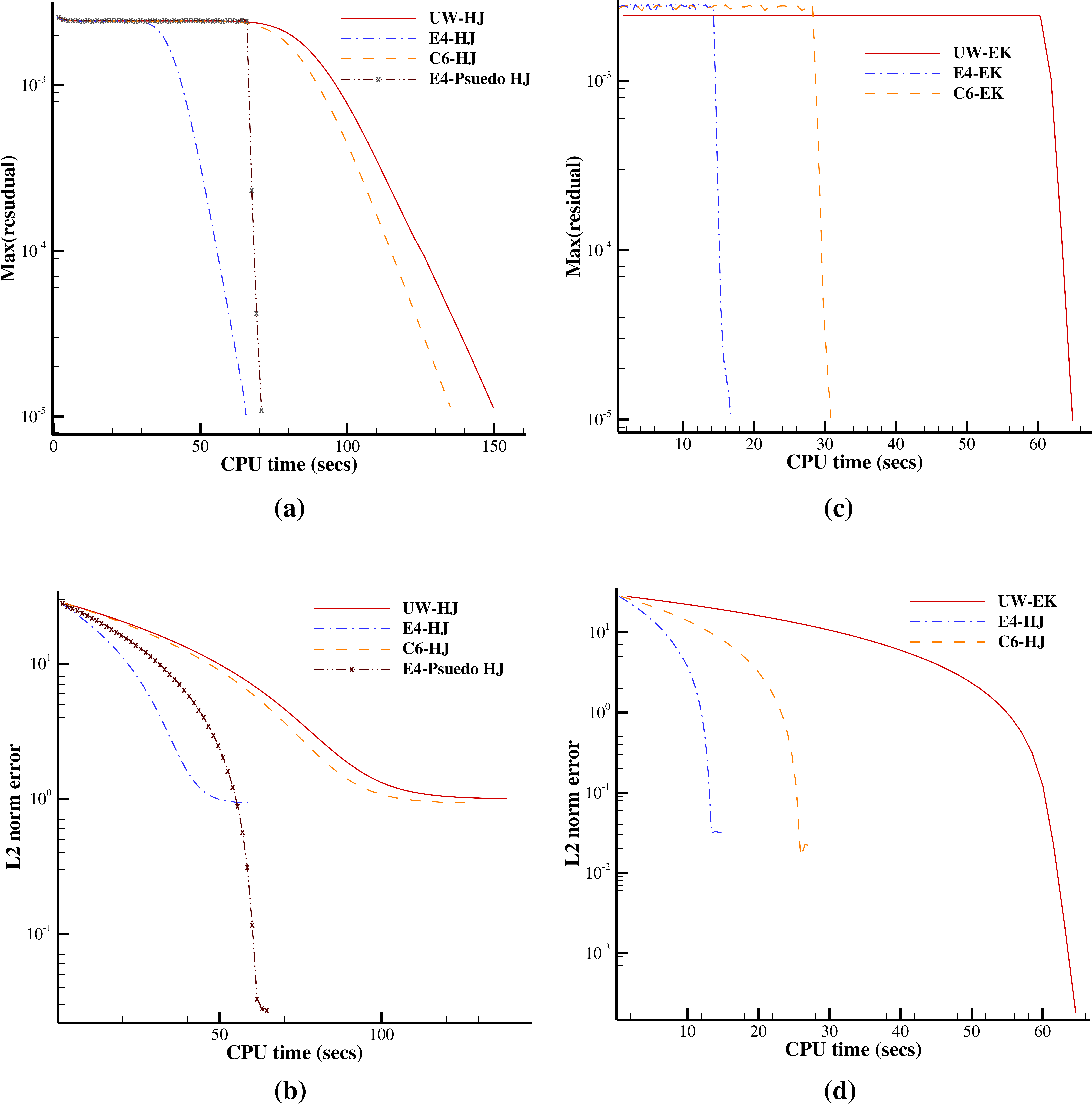}
\end{center}
\caption{Convergence histories of maximum residual (a,c) and $L^2$ norm error (b,d) for channel case using upwind and high-order schemes for Hamilton Jacobi (a,b) and Eikonal (c,d) equations.}
\label{converge} 
\end{figure}

\begin{figure}[h!]
\begin{center}
\includegraphics[width=150mm]{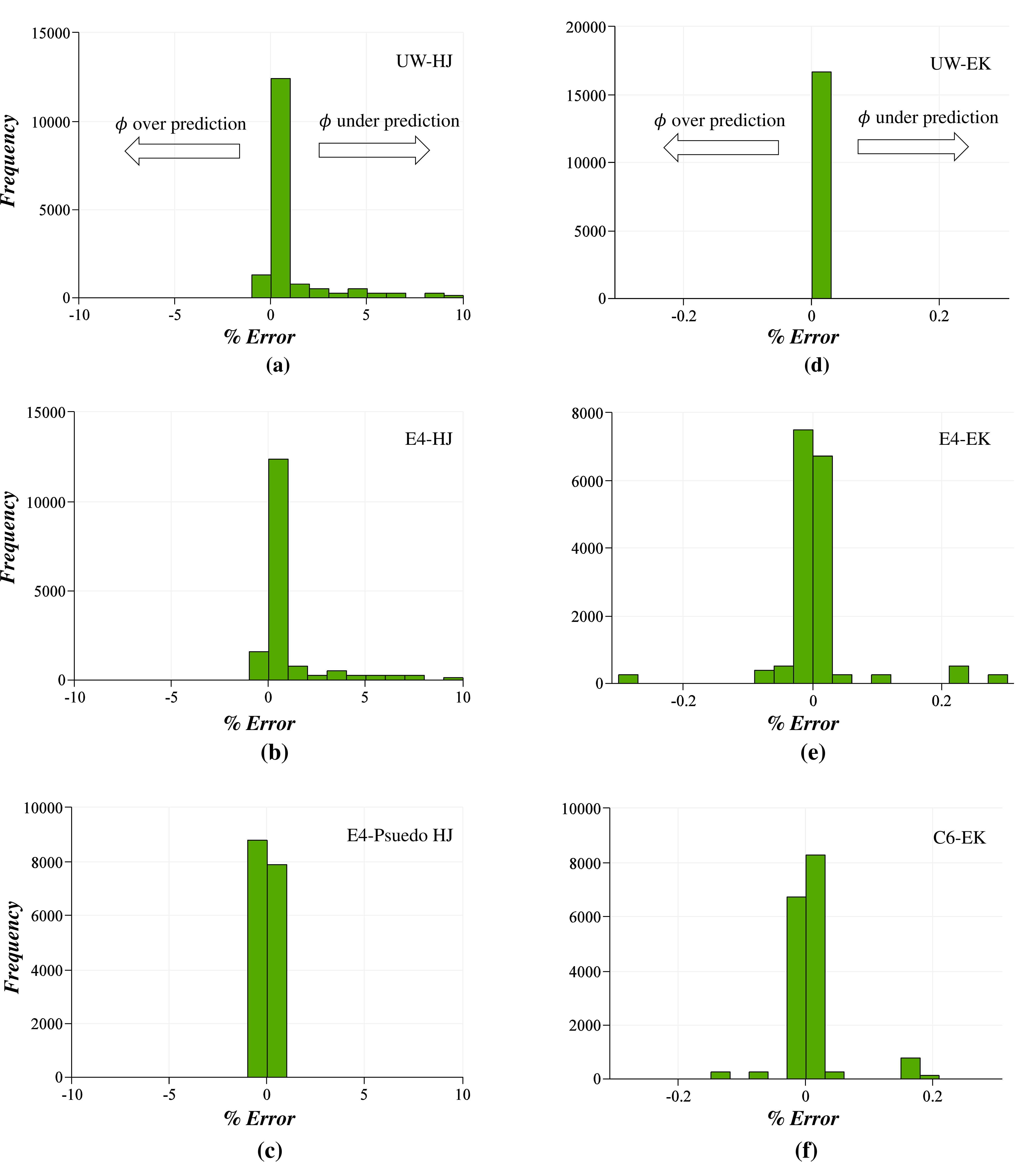}
\end{center}
\caption{Histograms of percentage error in wall distance estimates for channel case corresponding to Hamilton-Jacobi and Eikonal equations computed using upwind and high order schemes. The frequency on y-axis refers to the total number of solution points in the entire computational domain that constitute the percentage error range represented on the x-axis.}
\label{hist} 
\end{figure}

To further demonstrate the speed up using central-schemes, Figures \ref{hist}(a,c) show the convergence of the maximum residual given by Eq. \ref{USeq4} with CPU time for the channel case. In addition, Figures \ref{hist}(b,d) compare the convergence of L2 norm error (see Eq. \ref{l2norm}) with time for HJ and Eikonal schemes. From the plots, it is clear that the computational time associated with the upwind schemes is generally higher when compared to the central schemes. In particular, the explicit schemes offer more computational benefit and are almost $2\times$ faster than the compact/upwind schemes to converge to the same level of accuracy. This is attributed to the repeated computation of the wave front direction in upwind schemes. The effect on accuracy with different schemes is also minimal while solving the HJ equations. UW schemes are certainly suitable for the Eikonal based approach in terms of accuracy, as they can converge to lower values of L2 norm error (see Figure \ref{hist}(d)). However, as noted earlier, the compact (C4/C6) and Explicit schemes (E2/E4) are $\approx$ 2.3 to 3.5 times faster than the UW schemes respectively.

Following Tucker \cite{Tucker}, the error histograms for the HJ, Pseudo-HJ and Eikonal equations are estimated. The percentage errors are estimated as $(d_{act}-d)/d_{act} \times 100 \%$ where $d$ and $d_{act}$ are the predicted and true wall-distances. Figure \ref{hist} compares the histogram for different schemes. Negative errors represent over-prediction of wall distance w.r.t the true value and positive errors represent under-prediction. With HJ based approach, the scatter in the error is within $0-10\%$ with a positive bias (Fig. \ref{hist}(a,b)). This bias is attributed to the under-prediction of wall-distance away from the walls (see Fig. \ref{HJcases}). Figure \ref{hist}(c) shows a significant reduction in the scatter of the error using Pseudo-HJ. On the other hand, the error with Eikonal is within $\pm 0.2\%$ and is symmetrical than the HJ based approach (Fig. \ref{hist} (d,e,f)). Although the relative error magnitude is lower for all the schemes, the scatter in the error is much lower for UW-EK compared to the central-schemes.


\subsection{Effect of wall curvature and Nakanishi's modification}

The Laplacian term in H-J equation is dependent on the wall curvature and can assume very high or low values near convex and concave regions of the domain respectively. As a result, the H-J equation largely deviates from the Eikonal equation which results in high wall distance errors. Mathematically, $|\nabla^{2} \phi| \gg 0$ for convex regions. Hence, the RHS of the H-J equation $|\nabla \phi|=\sqrt{1+\Gamma \nabla^{2} \phi}$ will be much more greater than 1 in highly curved regions, there by significantly deviating from the Eikonal equation. For example, this effect can be seen at the airfoil leading and trailing edges. To counter this, Nakanishi \cite{Japanese} has proposed modifications to the original H-J equation. A curvature ($\kappa$) based correction is introduced into the formulation which can eliminate the influence of convex wall curvature on wall distance solution. The modified form reads as follows.

\begin{equation}
\textbf{U} \cdot \nabla \phi=1+(\epsilon \phi+v)\left\{\nabla^{2} \phi-MAX (0,|\nabla \phi| \kappa)\right\}
\end{equation}
\noindent The expressions for $\nu$ (which is introduced to enhance stability) and $\kappa$ are as follows: 

\begin{equation*}
	v=0.001(1-|\nabla \phi|)^{2}
\end{equation*}
\begin{equation*}
\kappa =\nabla \cdot \boldsymbol{n}, \quad \boldsymbol{n}=\frac{\nabla \phi}{|\nabla \phi| + \varepsilon_{0}}
\end{equation*}

\begin{figure}[h!]
\begin{center}
\includegraphics[scale=0.7]{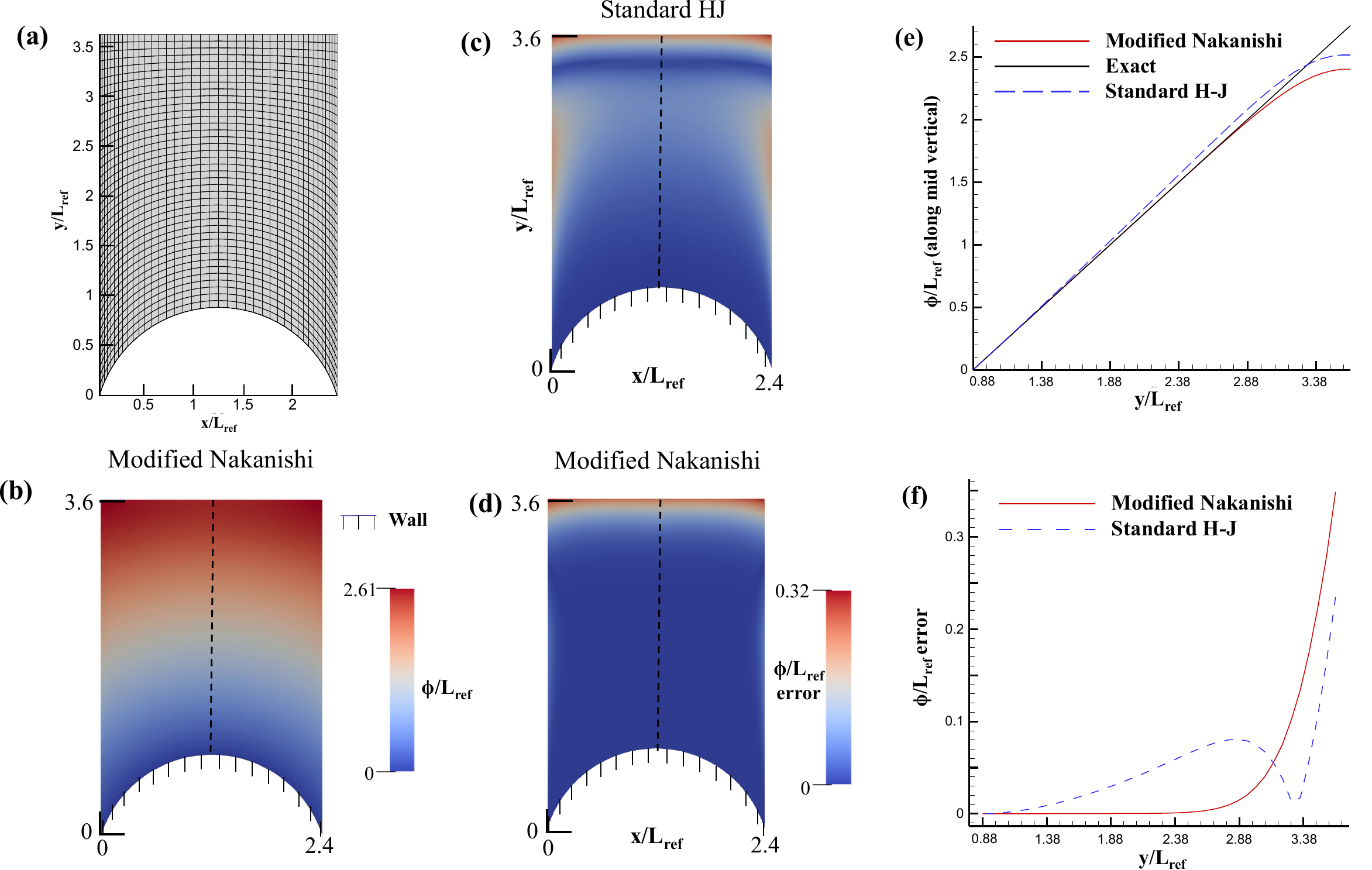}
\end{center}
\caption{(a) Computational grid used for the circular bump case;  (b) Wall distance contours ($\phi/L_{ref}$); Contours of wall distance error ($\phi/L_{ref} error$) using (c) standard H-J and (d) Modified Nakanishi approach; Comparison of (e) wall distance and (b) wall distance error obtained along the mid vertical of the computational domain. Simulations use E4 scheme with $\alpha_{f}$ = 0.495.}
\label{curvature2} 
\end{figure}


In the absence of filtering, the simulations with this formulation using E2 and E4 schemes are observed to be unstable due to large dispersion errors. To counter this, the value of $\epsilon$ can be increased, say to $1$, as proposed by Nakanishi \cite{Japanese}. However, higher values of $\epsilon$ (a) reduce the wall distance accuracy away from the walls and (b) reduce the stable pseudo time-step ($\Delta t^{*}$) thereby increasing the simulation time. To account for these accuracy and convergence issues, we have reduced the value of $\epsilon$ to 0.2 and used filtering (with $\alpha_{f} = 0.495$). We refer this approach as `Modified Nakanishi' in the plots.

Figure \ref{curvature2} shows the computational domain for the `circular bump' test case along with the contours of wall distance and absolute errors obtained using the standard H-J and modified Nakanishi approaches. From the wall distance and error plots shown in figures \ref{curvature2} (e) and (f), the effect of Nakanishi's formulation in mitigating the wall distance error can be clearly appreciated. The wall distance deviates from the exact solution using standard H-J approach even closer to the walls. On the other hand, the modified Nakanishi solution rectifies these curvature related issues and predicts accurate wall distances in most of the domain, particularly near the walls. It should be noted that Nakanishi's approach is recommendable for the applications where accurate wall distances are required at all the locations in the domain (CAD/automated meshing/computational interface for overset grids/acoustic ray tracing,etc). However, in the context of turbulence modelling, the Laplacian operator in the original formulation by Tucker et al. \cite{AR} has been deliberately tuned to exaggerate the wall-distance, $\phi>d_{true}$, near the fine convex features (thin wires) or convex corners to remedy the excessive turbulence dissipation. On the other hand, through $\phi< d_{true}$ at a concave corner, the Laplacian term naturally accounts for the damping effects due to extra walls. Similar benefit is also observed while using the Poisson-equation based approach for computing wall-distances for turbulence modelling \cite{Tucker}.


\subsection{Unsteady cases}

In this section, we discuss the results from the unsteady test cases like `piston cylinder arrangement', `bouncing cube' and `propellant grain burnback in solid rocket motor' using the Eikonal and H-J approach with E4 scheme are presented here. The computational set-up, grid and boundary conditions for these cases are described in Section \ref{casessec}.

\begin{figure}[h!]
\begin{center}
\includegraphics[width=170mm]{Images/unsteady1.pdf}
\end{center}
\caption{(a) Temporal evolution of wall distance field inside the piston-cylinder arrangement (case-7) computed using H-J equation (with epsilon = 0.15) (b) Comparison of wall-distance evolution at the probe location shown in the first snapshot with the exact solution.}
\label{unsteady1} 
\end{figure}

\begin{figure}[h!]
\begin{center}
\includegraphics[width=160mm]{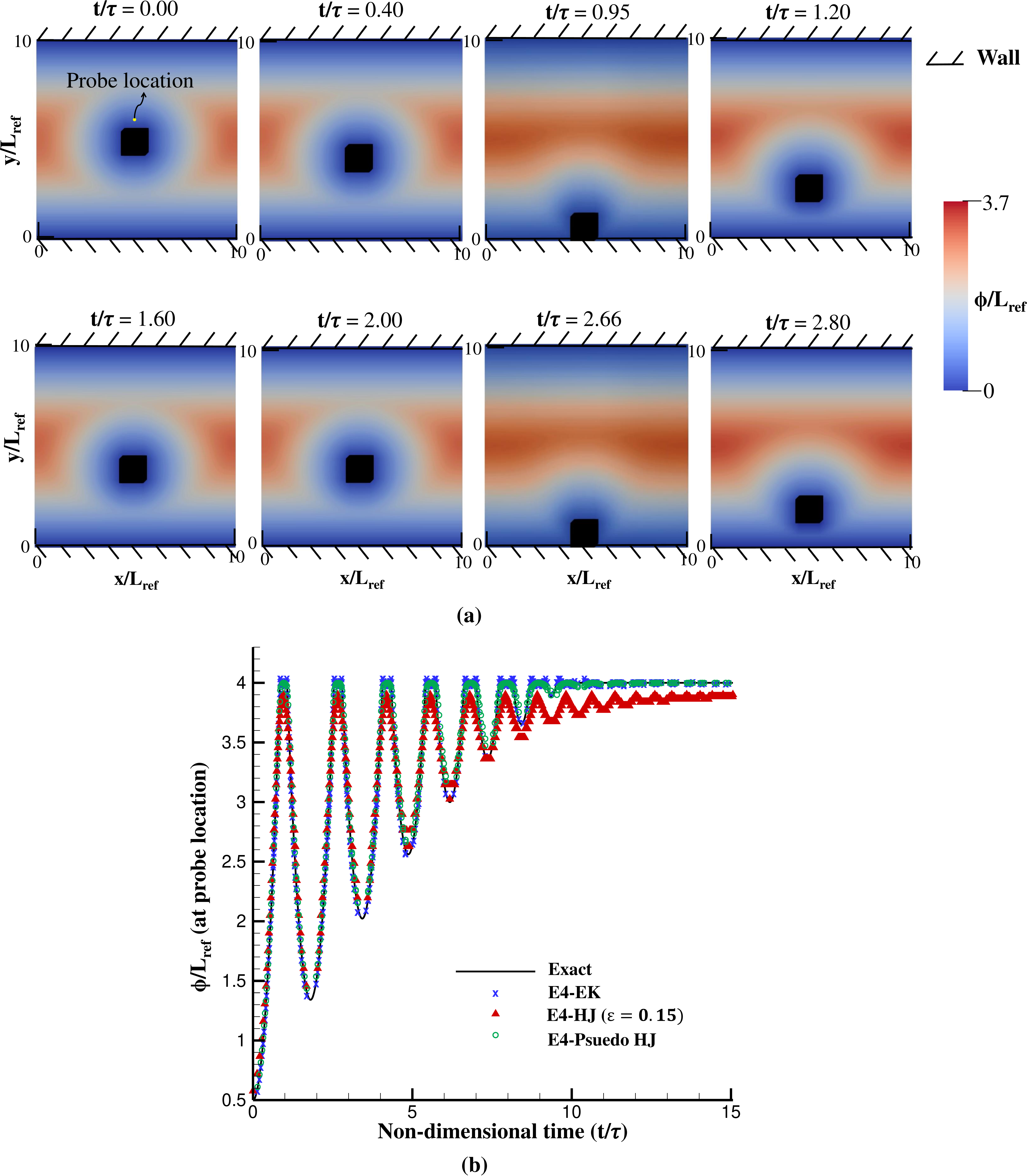}
\end{center}
\caption{(a) Temporal evolution of wall distance field around the bouncing cube (case-8) computed using H-J equation (with epsilon = 0.15) (b) Comparison of wall-distance evolution with the exact solution at the probe location shown in the first snapshot.}
\label{unsteady2} 
\end{figure}

\begin{figure}[h!]
\begin{center}
\includegraphics[width=160mm]{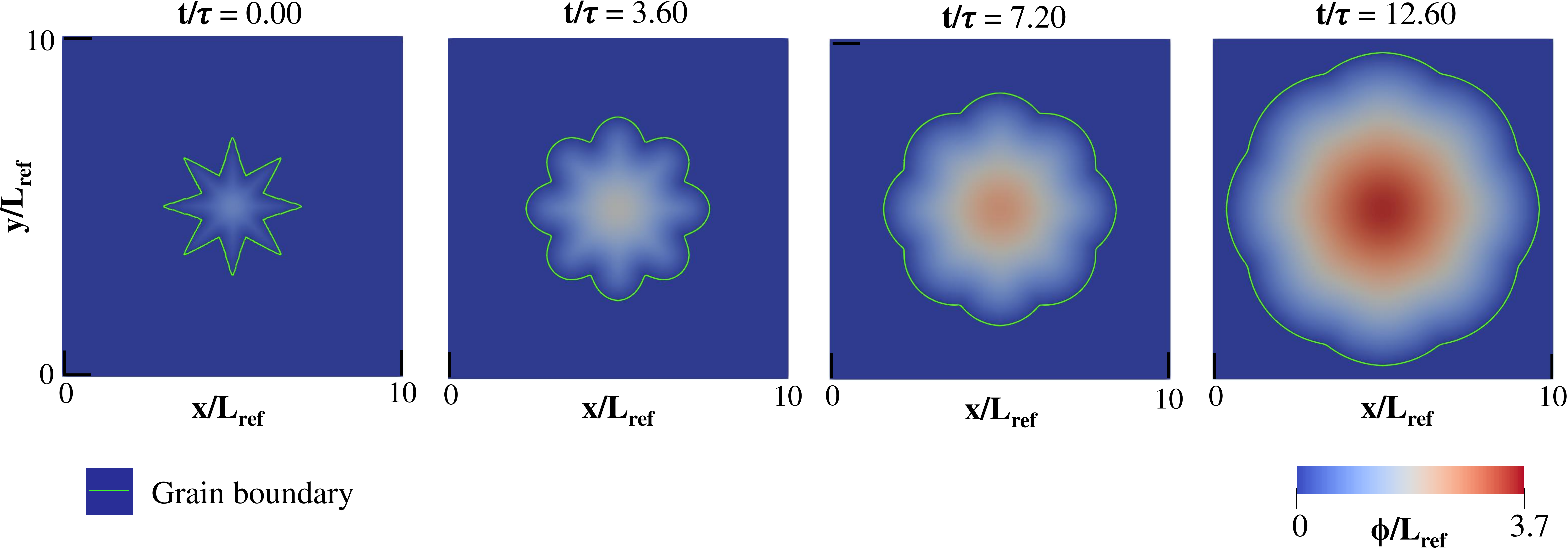}
\end{center}
\caption{Temporal evolution of wall distance field inside the combustion chamber of a solid rocket motor (with a 2-D eight legged star like propellant grain in it) computed using H-J equation.}
\label{unsteady3} 
\end{figure}

\noindent \textit{Canonical Cases:} Figures \ref{unsteady1}(a) and \ref{unsteady2}(a) show the temporal variation of the wall-distance field for the `piston-cylinder arrangement' and `bouncing cube' respectively. The conventional method of computing wall distances using search methods are particularly expensive in these kind of scenarios where the walls are moving. Soon after the first physical time-step, we have observed a drastic reduction in the simulation time for the subsequent time steps. This is attributed to the fact that the solution at the previous physical time step provides a close enough initial wall-distance estimate for the new physical time step thereby accelerating convergence. Figures \ref{unsteady1}(b) and \ref{unsteady2}(b) compare the temporal variation of wall-distance against the exact solution recorded at probes marked in Figures \ref{unsteady1}(a) and \ref{unsteady2}(a). The predictions of both Eikonal and H-J solutions are in good agreement with the analytical value. For the bouncing cube case, the cube looses its energy by 20\% every time it bounces off the ground, until it settles on the bottom surface. Hence, in Figure \ref{unsteady2}(b), an oscillatory convergence of the wall-distance to a steady value with increasing time is observable. Under the steady state, the probe is at a location which is farthest from all the walls. It has been demonstrated in Fig. \ref{unsteady2} that standard H-J equation is less accurate away from the walls. At $t/\tau$ = 10-15, a notable error between the analytical and predicted wall distance persists due to the Laplacian operator in the standard H-J equation. It is also apparent from Figures \ref{unsteady1}(b) and \ref{unsteady2}(b) that the results using the new H-J formulation with LAD are much more accurate than the standard H-J and are in encouraging agreement with the Eikonal and exact solutions.

\noindent \textit{Burnback of Propellant Grain:} One of the critical applications of unsteady wall-distances is in Solid propellant Rocket Motors (SRM), where the thrust-time or pressure-time characteristics of SRM has to be determined as a part of the design process. In SRM, a propellant grain of a given cross-sectional shape (star/tubular/double anchor/dendrite,etc) is burnt in the combustion chamber. For example, Fig. \ref{unsteady3} shows a 8 legged star shaped propellant grain burning radially outward. The equilibrium pressure generated inside the combustion chamber, $P_c$, during the burning process can be obtained by equating the rate at which hot gases are generated from the propellant surface (of instantaneous surface area $S_b$) to the rate at which hot gases leave the nozzle downstream (of throat area $A^*$). It is approximately given as:

\begin{equation}
    P_c = \left(aC^*(\rho_p-\rho_0)\frac{S_b}{A^*}\right)^\frac{1}{1-n}
    \label{pchamb}
\end{equation}

Here, $C^*$ is the characteristic velocity which is primarily a function of combustion chamber properties, $\rho_p$ is the propellant density and $\rho_0$ is the instantaneous density of the hot gases. $a$ and $n$ are empirical constants (see \cite{hill} for further details). It is evident from Eq. \ref{pchamb} that the instantaneous pressure $P_c$ within the SRM is a direct function of the instantaneous burning surface area of the propellant grain, $S_b$. Determining the temporal change in the shape of the propellant grain (and hence $S_b$) is crucial to accurately estimate the corresponding pressure-time characteristic. A level-set equation given below can be solved to determine the instantaneous shape of the grain. 

\begin{equation}
        \frac{\partial \phi_s}{\partial t} + {F} |{\nabla \phi_s}|=1
    \label{levelset}
\end{equation}

Here, $\phi_s$ is the signed wall-distance and $F$ is the regression speed of the grain boundary along the local normal. A zero level-set of the signed wall distance (iso-surface of $\phi_s$=0) represents the grain interface at that specific time instant. The gradient of signed wall distance is set to zero at all the other domain boundaries. Without loss of generality, we demonstrate the evolution of a grain boundary for a constant regression speed of $F = 0.3$. The effect of varying regression speed, as discussed in Wei et al. \cite{wei}, can also be accommodated in the current framework. The solution to this level-set equation is obtained in similar lines to that discussed in Sections \ref{geq} and \ref{spatial_disc}. Firstly, $\phi_{s,\xi}$, $\phi_{s,\eta}$ and $\phi_{s,\zeta}$ are estimated using the upwind schemes described in Eqns. \ref{UW3}-\ref{UW5}. Subsequently, $\phi_{s,x}$, $\phi_{s,y}$ and $\phi_{s,z}$ are estimated using the corresponding metric terms. Finally, $|{\nabla \phi_s}|$ in Eq. \ref{levelset} is determined as follows:

\begin{equation}
        |{\nabla \phi_s}| = (\phi_{s,x}^2+\phi_{s,y}^2+\phi_{s,z}^2)^{0.5}
\end{equation}

\begin{figure}[h!]
\begin{center}
\includegraphics[width=160mm]{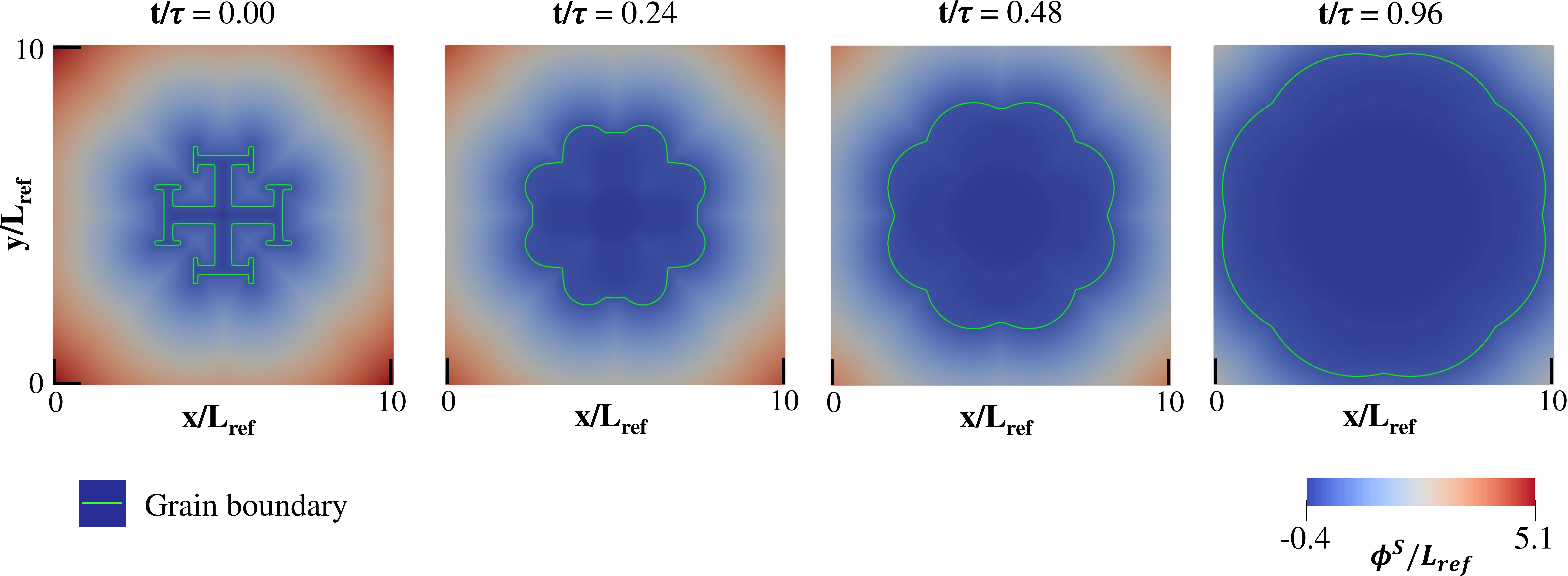}
\end{center}
\caption{Temporal evolution of signed wall distance field for a dendrite grain configuration. Green lines show the evolution of grain boundary.}
\label{unsteady4} 
\end{figure}

The final discretized equation is integrated in time using the standard explicit RK4 time-integration scheme. Figure \ref{unsteady3} shows the temporal evolution of a dendrite grain configuration. Figure \ref{unsteady4} shows the unsteady burning of a dendrite grain at different time instants. Contours show the signed wall-distance field and the lines show the grain boundary corresponding to a zero level set. The burning is radially outward and progressive. The instantaneous burning surface area of the propellant grain, $S_b$, increases with time and the sharp corners gradually smoothen out. Beyond $t/\tau \approx 0.96$, $S_b$ decreases during the web burnout phase.

\begin{figure}[h!]
\begin{center}
\includegraphics[width=165mm]{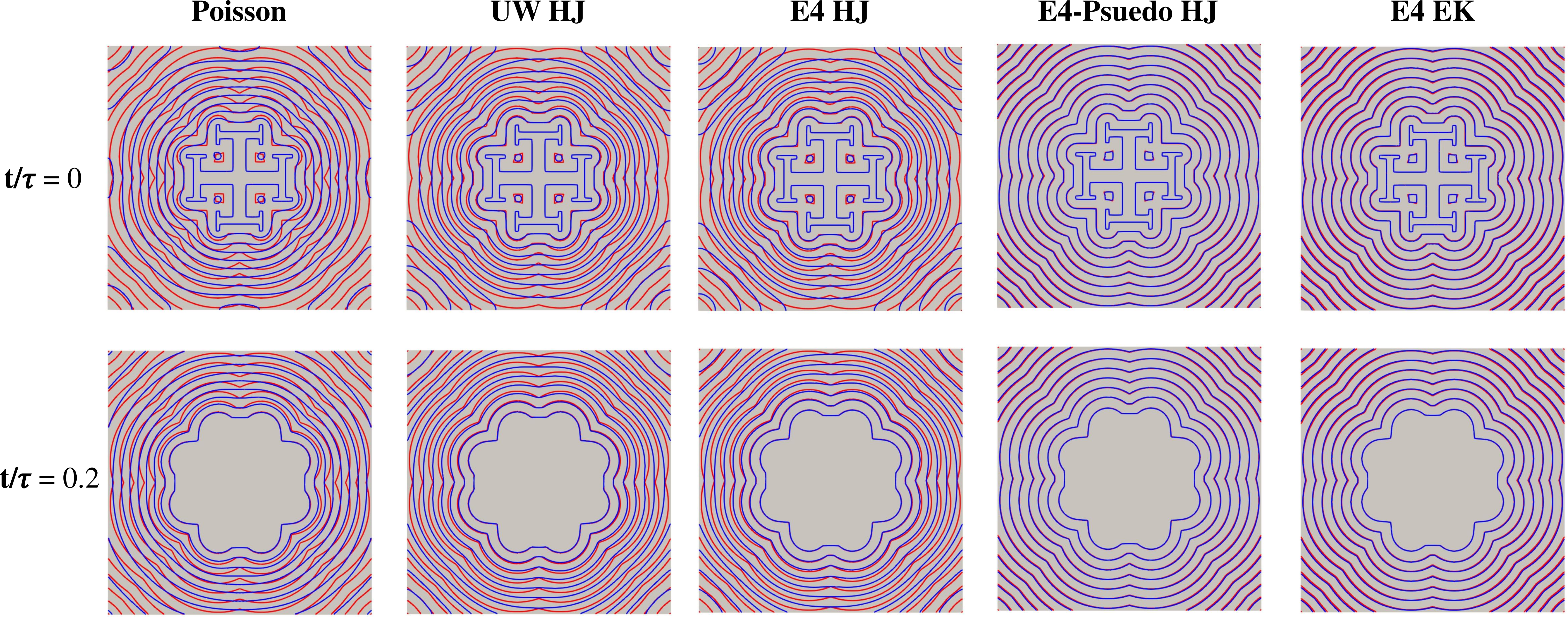}
\end{center}
\caption{Line contours of signed wall distance function at two time instances. Red lines correspond to the reference case where wall-distance field is initialized solving Eikonal equation using upwind schemes, and blue lines show wall-distance initialized using other schemes (Poisson/HJ/HJ LAD/ E4 Eikonal)}
\label{unsteady5} 
\end{figure}

\begin{figure}[h!]
\begin{center}
\includegraphics[width=175mm]{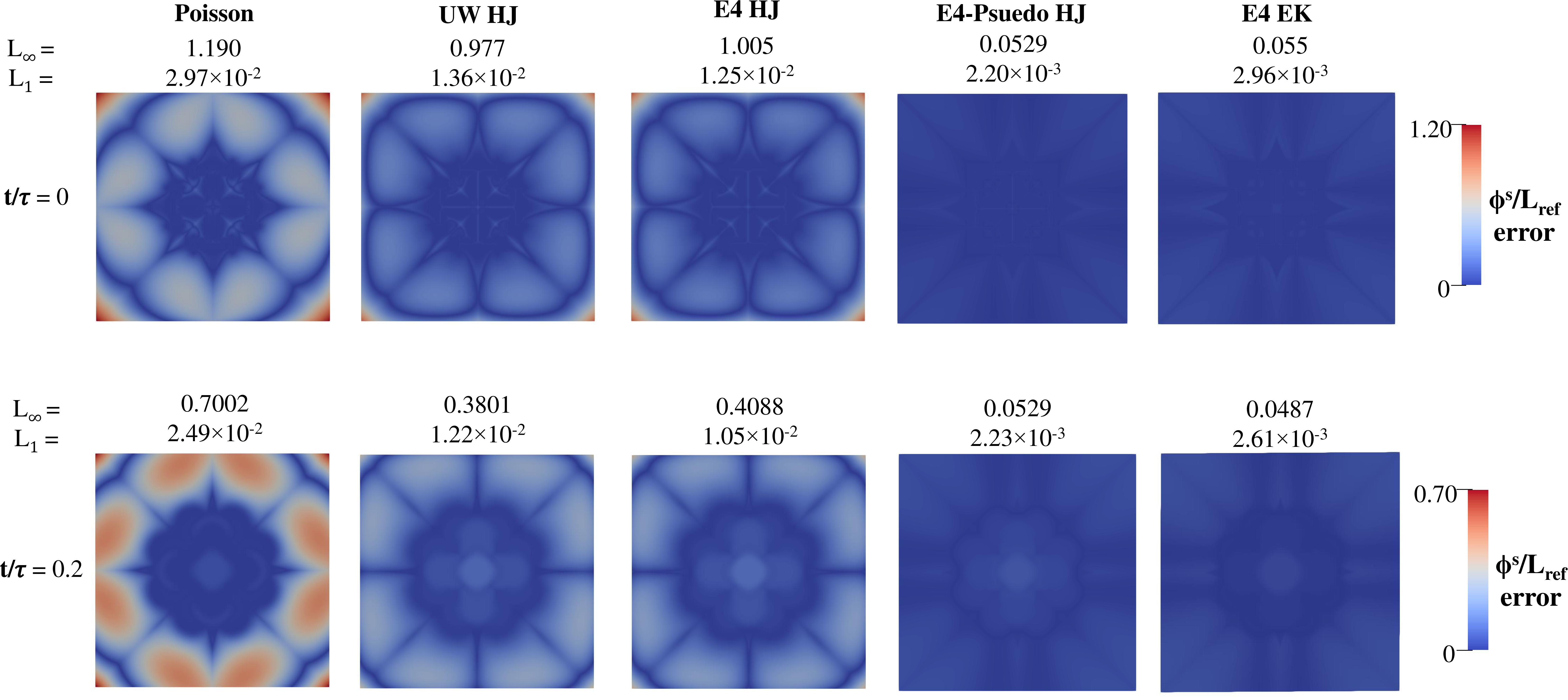}
\end{center}
\caption{Absolute error contours of the signed wall distance function w.r.t the UW EK based initialization.}
\label{unsteady6} 
\end{figure}

To initialize the aforementioned level-set procedure at $t/\tau = 0$, a signed wall-distance field, $\phi_s$, must be computed for the initial dendrite grain configuration using any of the approaches (Poisson/HJ/HJ LAD/Eikonal) discussed in the previous sections. Subsequently, the level-set equation (Eq. \ref{levelset}) can be solved to track the evolution of the grain boundary. Figure \ref{unsteady5} compares the effect of different initialization procedures on the evolution of the level-set contours. For comparison, a wall-distance field initialized by solving the Eikonal equation using upwind schemes is overlaid. Both the Eikonal and HJ-LAD based initialization, obtained using explicit 4th order schemes, show an encouraging agreement against the reference. On the other hand, discrepancies are evident in the level-set contours when the wall-distance field is initialized either using the baseline HJ or Poisson equations due to their inaccurate predictions away from the wall. Figure \ref{unsteady6} compares the corresponding contours of the absolute error for different initialization strategies. Clearly, the $L_1$ and $L_\infty$ errors using Eikonal and LAD based HJ schemes are an order of magnitude lower than the level-set contours predicted using Poisson/baseline HJ based initialization.

\begin{figure}[h!]
\begin{center}
\includegraphics[width=145mm]{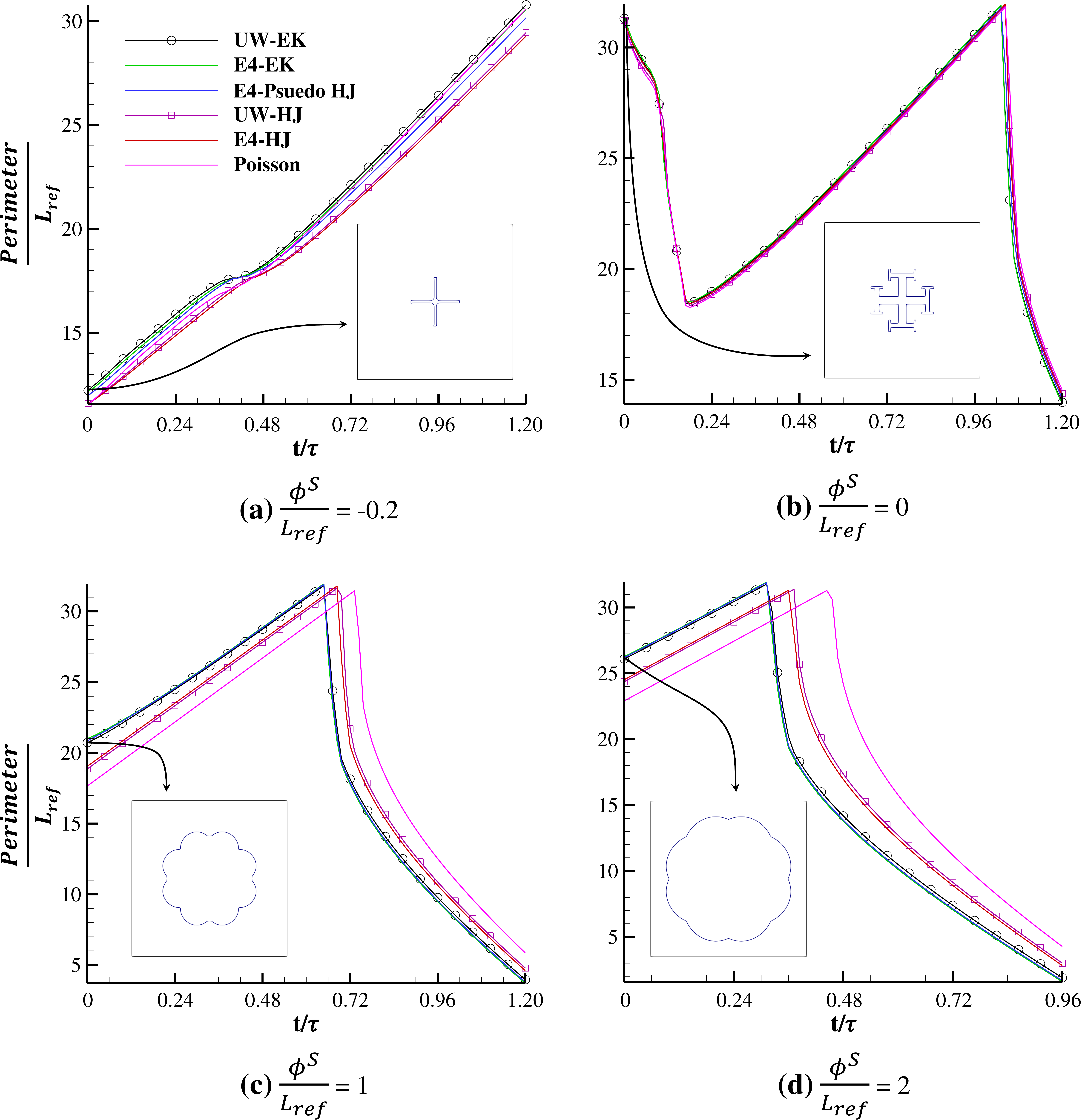}
\end{center}
\caption{Comparison of the temporal evolution of the grain perimeter with different initializations of the signed wall-distance field (obtained from different wall-distance approaches). (a-d) correspond to the iso-lines of different signed wall distance levels, $\phi^S/L_{ref}$, as shown in the inset plots at $t/\tau=0$.}
\label{unsteady7} 
\end{figure}

It should however be noted that despite these differences in the level-set evolution, the grain-boundary can be accurately tracked using any of the aforementioned wall-distance based approaches as these are accurate close to the wall. To demonstrate this, iso-lines corresponding to different level-set values have been tracked in time. The grain is two-dimensional (extruding into the page) and hence the perimeter of level-set iso-lines is proportional to the surface area. Recall from Eq. \ref{pchamb} that the grain surface area in turn determines the chamber pressure. Figures \ref{unsteady7} (b,c,d) show the temporal evolution of the perimeter for the level-sets of $\phi^s/L_{ref}$ = 0, 1 and 2. Of these, $\phi^s/L_{ref}$ = 0 represents the true grain boundary. Corresponding level-set iso-lines at $t/\tau = 0$ are shown in the inset plots for reference. Clearly, the wall-distance initialization procedure has little impact on the temporal evolution of the grain surface area. The typical two-peak evolution observed in pressure-time characteristic of a dendrite grain is captured. An initial drop in the surface area due to the burning of the dendrite projections is evident up to $t/\tau \approx 0.23$. This is followed by an increase in the surface area up to $t/\tau \approx 0.97$, beyond which it drops during the web burnout phase. Discrepancies due to the initial wall-distance field are observable for $\phi^s/L_{ref}$ = 1 and 2; farther the level set higher the error. Consistent with the observations made earlier, predictions using E4-EK and E4-LAD are as accurate as UW-EK. 

Apart from assisting in tracking the grain boundary, wall-distance field is also used in the turbulence models to predict the flow within the increasing chamber volume where the hot gases are generated (see figure \ref{unsteady3} which shows wall-distance field enclosed within the grain boundary). Hence, the perimeter of a level set corresponding to $\phi^s/L_{ref}$ = -0.2 is also tracked and is shown in \ref{unsteady7} (a) demonstrating the sensitivity of the solution to the initial wall-distance field. It is worth noting that in typical CFD applications, it is sufficient to compute accurate wall distances close to the wall. On this front, even a Poisson equation yields acceptable results as illustrated in Figure \ref{baseline1}(g). Nevertheless, accurate wall distances away from the wall are required in several other applications like evaluating the computational interface for overset grids \cite{nakahashi}, acoustic ray tracing inside jet mixing layers \cite{xia}, etc. The current study demonstrates the computational efficacy of using LAD based HJ methods or Eikonal schemes using high-order schemes with filtering to yield accurate results in such applications.

\textit{Recommendations:} The discussion so far has clearly brought out the benefits and disadvantages of different differential equation-based wall-distance based approaches (Poisson/Hamilton-Hacobi/Pseudo-HJ/Eikonal) using several spatial discretization schemes (Central schemes - explicit/compact and Upwind). The choice of a particular wall-distance approach depends on the application of interest. For applications where accurate wall-distances are required away from wall (CAD/automated meshing/computational interface for overset grids/acoustic ray tracing,etc), Eikonal/ LAD based HJ approach is preferable. On the other hand, standard HJ can be used for turbulence modelling applications as the Laplacian operator is tuned to exaggerate the wall-distance, $\phi>d_{true}$, near the fine convex features. We recommend explicit $4^{th}$ order (E4) central difference scheme for solving either of these governing equations as it offers benefits both in terms of speed and accuracy.

\section{Conclusions} \label{conc}

In this study, we have explored the efficacy of several differential equation based wall distance approaches (Poisson/Hamilton-Jacobi, HJ/Pseudo HJ/Eikonal) using both upwind and high order central difference (Explicit/Compact) schemes for spatial discretization. Firstly, the baseline solver which employs the commonly used upwind spatial discretization scheme is developed and validated on the canonical test cases. The computational benefit of using central difference schemes (E2, E4, C4 and C6) for wall-distance solvers, both in terms of computational time and accuracy, has been demonstrated. In addition, a modified pseudo Hamilton-Jacobi (H-J) formulation based on Localized Artificial Diffusivity (LAD) approach has been proposed. A modified curvature correction has been implemented into the HJ solver to account for the error due to concave/convex wall curvatures. Using both the canonical and the unsteady test cases where the wall-distance evolves with time, the new pseudo HJ approach is demonstrated to be as accurate as the Eikonal approach. The unsteady test cases include `piston-cylinder arrangement', `bouncing cube' and wall-distance evolution inside the combustion chamber of a solid rocket motor due to the burning of a `star grain propellant'.

While solving the Hamilton-Jacobi equation, the high-order central difference schemes performed approximately $1.4-2.8$ times faster than the upwind schemes with a marginal improvement in the solution accuracy. The HJ based on LAD scheme produced results which are of comparable accuracy to the Eikonal solution. It's solution accuracy is almost one order higher and the calculations are $\approx$ 1.5 times faster than the baseline HJ solver using upwind schemes. On the other hand, while solving the nonlinear hyperbolic Eikonal equation, the global error from high-order schemes is found to be larger when compared to the upwind schemes. Due to the lack of Laplacian operator in the Eikonal equation, the central schemes suffered with some dispersion errors despite filtering.



\begin{thebibliography}{99}

\bibitem{Tucker}Tucker, P. G., \textit{Differential equation-based wall distance computation for DES and RANS}, Journal of computational physics. Volume 190, Issue 1, 229-248 (2003)

\bibitem{AR}Tucker, P. G., Rumsey, C. L., Spalart, P. R., Bartels, R. E., and Biedron, R. T., \textit{Computations of wall distances based on differential equations}, AIAA journal, 43(3), 539-549. (2005)

\bibitem{JL}Richard J. Jefferson-Loveday, V. Nagabhushana Rao, James C. Tyacke and Paul G. Tucker, \textit{High-order detached eddy simulation, zonal LES and URANS of cavity and labyrinth seal flows}, Int. J. Numer. Meth. Fluids. (2013)

\bibitem{RS_AST}Tucker, P. G., \textit{Hybrid Hamilton–Jacobi–Poisson wall distance function model},  Computers \& fluids, 44(1), 130-142. (2011)

\bibitem{skews}Jing-lei XU, Chao YAN, Jing-jing FAN. \textit{Computations of wall distances by solving a transport equation}, Appl. Math. Mech. -Engl. Ed., 32(2), 141–150. (2011)

\bibitem{ISSW}Anthony Bouchard, \textit{Wall distance evaluation via Eikonal solver for RANS application}, Masters Thisis, UNIVERSITÉ DE MONTRÉAL. (2017)

\bibitem{visbal}Miguel R. Visbal and Datta V. Gaitonde, \textit{High-Order Schemes for Navier-Stokes Equations: Algorithm and Implementation Into FDL3DI}, AFRL. (1998)

\bibitem{rich}Jefferson-Loveday, R. J., Tucker, P. G., Northall, J. D., Nagabhushana Rao, V. 
\textit{Differential equation specification of integral turbulence length scales}, Journal of turbomachinery, 135(3). (2013)

\bibitem{First_EK_paper}E. Fares  W. Schröder, \textit{A differential equation for approximate wall distance}, Int. J. Numer. Meth. Fluids, Volume39, Issue8, (2002)

\bibitem{raothesis} Vadlamani, N. R., \textit{Numerical investigation of separated flows in low pressure turbines}, Doctoral dissertation, University of Cambridge. (2014)

\bibitem{SRM} Arnau Pons Lorente, \textit{Study of grain burnback and performance of solid rocket motors.}, Project report, Universitat Politècnica de Catalunya. (2013)

\bibitem{Turb_model} Spalart, P. R. and Allmaras, S. R. , \textit{One-Equation Turbulence Model for Aerodynamic Flows}, La Recherche Aerospatiale, 1 (1994), pp. 5–21.

\bibitem{search1} Sebastian G., Charbel F., \textit{Fast computation of the wall distance in unsteady Eulerian fluid‐structure computations}, Int. J. Numer. Meth. Fluids, Volume 89, Issue 4-5, 143-161 (2019)

\bibitem{search2} Beatrice Roget, Jayanarayanan Sitaraman, \textit{Wall distance search algorithm using voxelized marching spheres}, J. Comp Physics, Volume 241, 15 May 2013, Pages 76-94 (2013)

\bibitem{CAD_paper} Les A. Piegl, Wayne Tiller, \textit{Algorithm for finding all k nearest neighbors}, Journal of Computer Aided Design, Volume 34, Issue 2, 82-172 (2002)

\bibitem{Auto_mesh} Andrea Milli, Shahrokh Shahpar, \textit{PADRAM: Parametric Design and Rapid Meshing System for Complex Turbomachinery Configurations}, ASME Turbo Expo, Copenhagen, Denmark (2012)

\bibitem{COMP_SQ1} Achu, Shankar, and Nagabhushana Rao Vadlamani, \textit{ Entropically Damped Artificial Compressibility Solver Using Higher Order Finite Difference Schemes on Curvilinear and Deforming Meshes.} In AIAA Scitech 2021 Forum, p. 0634. (2021)

\bibitem{COMP_SQ2} Vadlamani, Nagabhushana Rao, Paul G. Tucker, and Paul Durbin, \textit{Distributed roughness effects on transitional and turbulent boundary layers}, Flow, Turbulence and Combustion 100, no. 3 p. 627-649. (2018)

\bibitem{COMP_SQ3} Lin, Y., Vadlamani, R., Savill, M. and Tucker, P., \textit{Wall-resolved large eddy simulation for aeroengine aeroacoustic investigation}, The Aeronautical Journal, 121 (1242), pp.1032-1050. (2017)

\bibitem{LAD} S. Kawai, S.K. Lele, \textit{Localized artificial diffusivity scheme for discontinuity capturing
on curvilinear meshes}, Journal of Computational Physics 227 (2008) 9498–9526

\bibitem{Poisson_first} Spalding DB. Calculation of turbulent heat transfer in cluttered spaces. Presented at the 10th international heat transfer conference, Brighton, UK; (1994).

\bibitem{Japanese} Tameo Nakanishi, \textit{A Modification on Hamilton-Jacobi (HJ) Equation for Computing Wall Distances}, Trans. Japan Soc. Aero. Space Sci., Vol. 49, No. 163, pp. 55–57, (2006)

\bibitem{srm} Yildirim, Cengizhan and Aksel, Haluk, \textit{Numerical simulation of the grain burnback in solid propellant rocket motor}, In 41st AIAA/ASME/SAE/ASEE Joint Propulsion Conference \& Exhibit, p. 4160 (2005).

\bibitem{hill} Hill, Philip G., and Carl R. Peterson. \textit{Mechanics and thermodynamics of propulsion}, Reading (1992).

\bibitem{wei} Wei, R., Bao, F., Liu, Y., \& Hui, W, \textit{Robust three-dimensional level-set method for evolving fronts on complex unstructured meshes.} Mathematical Problems in Engineering, (2018).

\bibitem{nakahashi} Nakahashi K, Togashi F, \textit{Intergrid-boundary definition method for overset unstructured grid approach.} AIAA J 2000;38(11):2077–84.

\bibitem{xia} Xia, H., Tucker, P. G., \& Dawes, W. N. \textit{ Level sets for CFD in aerospace engineering.} Progress in Aerospace Sciences, 46(7), 274-283 (2010).

\end{thebibliography}

\end{document}